\documentclass[
    reprint,
    preprintnumbers,
    superscriptaddress,
    nofootinbib,
    amsmath,amssymb,
    aps,
    prd,
    floatfix,
    longbibliography]{revtex4-2}
    
\usepackage{aas_macros}
\usepackage{amsmath,amssymb}
\usepackage{graphicx}
\usepackage{xcolor}
\usepackage{hyperref}
\usepackage{comment}

\hypersetup{
  breaklinks = true,
  colorlinks   = true, 
  urlcolor     = blue, 
  linkcolor    = blue, 
  citecolor   = blue 
}

\DeclareUnicodeCharacter{2212}{\ensuremath{-}}
\begin{document}

\title{Consistency between cosmological and standard siren observations in evolving dark energy}

\author{Macarena Lagos}
\affiliation{Institute of Astrophysics, Department of Physics and Astronomy, Universidad Andrés Bello, Santiago, Chile}
\email{macarena.lagos.u@unab.cl}
\author{William J. Wolf}
\affiliation{Department of Physics, University of Oxford, DWB, Keble Road, Oxford OX1 3RH, United Kingdom}
\email{william.wolf@physics.ox.ac.uk}
\date{\today}

\begin{abstract}
A scalar field non-minimally coupled to gravity can be the driver of cosmic acceleration. Such a non-minimal coupling (NMC)  can produce a
non-zero gravitational-wave (GW) friction function $\alpha_M(z)$, which
modifies the luminosity distance inferred from GW sources relative to its
electromagnetic counterpart. 
We use a particular NMC scalar-tensor model, that explains time-varying dark energy in good alignment with DESI, to predict $\alpha_M(z)$ and the
expected GW/EM luminosity-distance ratio $D_L^{\rm GW}/D_L^{\rm EM}$, and
map it onto two common parametrizations---the $c_M$ and the
$(\Xi_0,n)$ models.  We find $c_M = -0.5\pm 0.2$ and $\Xi_0 = 0.88\pm 0.05$,
$n=3.2\pm 0.3$, both consistent with GWTC-5 constraints at the $\lesssim 1\sigma$ level. This consistency is mostly driven by current large uncertainties in GW data, which lead to measurements consistent with both $\Lambda$CDM and the NMC model. 
By contrast, the dark energy constraints derived from analyzing cosmological data, under the common parametric scalar-tensor model using $\alpha_M(z)=c_M\Omega_\Lambda(z)/\Omega_{\Lambda 0}$ and the CPL parametrized equation of state $w_0w_a$, are in $2.2\sigma$ ($3.6\sigma$) tension with the NMC predictions for the $c_M$ ($w_0w_a$) parameter. We confirm that, in order to avoid cosmological instabilities, this parametrized model imposes strong implicit priors that are incompatible with physically-motivated scalar-tensor models when dark energy is dynamical.
\end{abstract}

\maketitle

\section{Introduction}
\label{sec:intro}

The accumulation of high-precision cosmological observations---baryon
acoustic oscillations (BAO) from DESI~\cite{DESI:2025zgx}, the cosmic microwave
background (CMB) from Planck and ACT \cite{Planck:2018vyg, ACT:2023kun, Planck:2019nip}, and Type~Ia supernovae (SNe) \cite{Scolnic:2021amr, DES:2024jxu, DES:2025sig, Rubin:2023jdq, Hoyt:2026fve}---has provided increasingly intriguing  evidence for \emph{dynamical} dark
energy, with an equation-of-state parameter that departs from $w=-1$ at
redshifts $z\lesssim 2$. This preference for dynamical dark energy is present at the $3.1\sigma$ level with only CMB and BAO data, and the full significance of the tension between $\Lambda$ and dynamical dark energy ranges between $2.8\sigma-4.2\sigma$ depending on the exact SNe data deployed in the analysis \cite{DESI:2025zgx}. 

This data is often interpreted through the lens of phenomenological parametrizations such as the
Chevallier-Polarski-Linder (CPL) \cite{Linder:2002et, Chevallier:2000qy} form $w(a)=w_0+w_a(1-a)$. While they can fit the data
well, they shed little light on the underlying microphysics \cite{Wolf:2023uno, Wolf:2024eph}. Besides $\Lambda$, the simplest and most natural way of modeling dark energy is with a dynamical scalar field $\varphi$ minimally coupled to gravity, known as quintessence \cite{Ratra:1987rm, Peebles:1987ek, Caldwell:1997ii, Ferreira:1997hj}. However, these simple realizations do not describe the data as well as the $(w_0, w_a)$ models \cite{Wolf:2024eph, DESI:2025fii, Payeur:2024dnq, Bhattacharya:2024hep} and the evidence for quintessence is marginal at best. 

These results, along with a curious, but persistent preference for dark energy that was apparently phantom in the recent past \cite{DESI:2025fii, Berti:2025phi}, have led to a wider exploration of more exotic scalar field proposals.
A physically motivated candidate is a scalar field $\varphi$ non-minimally
coupled to gravity through a term $\xi\varphi^2 R$ in the action.  Wolf et
al.~\cite{Wolf:2025jed} recently showed that such a non-minimally coupled (NMC)
scalar-tensor model can be strongly preferred over $\Lambda$CDM by the combination
of DESI DR2 BAO, CMB, and SNe cosmological data. This model reproduces or exceeds the ability of the $(w_0, w_a)$ models to fit the expansion history data across a wide range of statistical measures, in large part due to its ability to reproduce the phantom crossing of $w(a)$ favored by the data \cite{Wolf:2025jed, Wolf:2024stt, Ye:2024ywg}.

The non-minimal coupling has a further crucial implication for gravitational-wave (GW)
astronomy: it induces a time-varying effective
gravitational coupling on cosmological scales, quantified by the effective Planck mass running rate $\alpha_M(z) \equiv d\ln M_*^2/d\ln a$. This parameter generates a ``GW friction'' that modifies the
amplitude of GW signals, leading to a GW luminosity distance $D_L^{\rm GW}$
that differs from the electromagnetic luminosity distance $D_L^{\rm EM}$
\cite{Saltas:2014dha, Nishizawa:2017nef,Belgacem:2018lbp, Lagos:2019kds}. Further, amongst the numerous interactions that can produce a phantom crossing equation of state within the single scalar field paradigm\footnote{See coupled dark matter-dark energy models and Galileon models which can produce very similar expansion histories to the NMC model and likewise offer comparable descriptions of the cosmological data \cite{Khoury:2025txd, Gomez-Valent:2026ept, Chakraborty:2025syu, Tsujikawa:2025wca, Wolf:2025acj, Hallam:2026qsk, Khoury:2026svx, Naidoo:2026umv}.}, this GW effect is exclusively to theories which induce a time-varying Planck mass, meaning that this effect offers a unique cosmological probe of direct couplings to $R$. This effect is probed by multi-messenger standard sirens and by the population of binary black hole mergers observed by the LIGO-Virgo-KAGRA (LVK) collaboration. The latest LVK analysis of the fifth Gravitational-Wave Transient Catalog (GWTC-5)
\cite{LIGOScientific:2026uyd} constrains two phenomenological parametrizations of
GW propagation: (i) the $c_M$ model, in which
$\alpha_M(z) = c_M\,\Omega_\Lambda(z)/\Omega_{\Lambda 0}$ is assumed to evolve with the fractional dark energy density $\Omega_\Lambda(z)$ \cite{Lagos:2019kds}; (ii) the
$(\Xi_0,n)$ model, which directly parametrizes the distance ratio $D_L^{\rm GW}/D_L^{\rm EM}=\Xi_0+(1-\Xi_0)/(1+z)^n$ \cite{Belgacem:2018lbp}.

A parallel line of work has shown that the simple parametrizations used in
both CMB/LSS and GW analyses impose non-trivial theoretical priors on the
space of dark energy models.  Shah et al.~\cite{Shah:2025vnt} demonstrated that
the $\alpha_M \propto \Omega_{\Lambda}$ parametrization, ubiquitous in LSS
analyses, forces strong correlations between the dark energy background
($w_0, w_a$) and perturbation parameters, so that constraints derived within
this framework can be misleading when the true underlying model has a
different structure.  Indeed, Fig.~17 of Ref.~\cite{Shah:2025vnt} shows that
\emph{any} parametrization of perturbations within the Effective Field Theory of Dark Energy (EFTDE) \cite{Gubitosi:2012hu, Bellini:2014fua} disfavors the large
negative $w_a$ values preferred by DESI, whereas DESI data prefer precisely
such values---a direct manifestation of the implicit prior problem.

In this paper we address the following questions: are the GW-propagation
signatures predicted by the NMC model of Wolf et al.\ consistent with the
current GW data, and how do they compare to what is inferred from cosmological
analyses using the agnostic EFTDE parametrizations?  We answer this by:
(i) projecting the $\alpha_M(z)$ posterior of Wolf et al.\ onto the GW
observables $D_L^{\rm GW}/D_L^{\rm EM}$; (ii) fitting the standard $c_M$ and
$(\Xi_0,n)$ models to those projected posteriors; (iii) quantifying
the agreement between the NMC prediction and the GWTC-5 GW constraints; (iv) comparing the NMC model constraints to those coming from agnostic EFTDE parametrizations fitted with purely cosmological data.

Projecting the NMC predictions onto the GW observables, we find a present-day GW friction corresponding to $c_M = -0.5\pm 0.2$ and $(\Xi_0, n) = (0.88\pm 0.05,\, 3.2\pm 0.3)$, both consistent with the direct GWTC-5 constraints at the $\lesssim 1\sigma$ level. In contrast, the dark energy parameters ($w_0,w_a,c_M$) obtained from the most common EFTDE parametrization $c_M \propto \Omega_{\Lambda}$ fitted to purely cosmological data (CMB/LSS/SN) are in $\sim 2-4\sigma$ tension with those from the NMC model. We trace this issue to the fact that this EFTDE model restricts the parameter space of dark energy in order to avoid cosmological instabilities, in such a way that it naturally disfavors time-varying dark energy. 
This shows that this widely used parametrization imposes strong implicit priors that are incompatible with physically motivated scalar-tensor models with dynamical dark energy.

The paper is organized as follows. Section~\ref{sec:theory} summarizes the predictions for the viable NMC model we consider, and presents the results of fitting the two standard parametrizations for the GW friction effect. Section~\ref{sec:GW_agreement} presents our comparisons between GW data sets and the NMC cosmological analysis, including the impact that the non-zero value of $\alpha_M$ predicted by the NMC model can have on standard sirens constraints on $H_0$. Section \ref{sec:Shah_agreement} presents the comparison between the constraints coming from EFTDE parametrized cosmological models and the NMC model, using different cosmological data sets.
Finally, section~\ref{sec:discussion} discusses the implications of our results and future
directions. Throughout this paper we will quote results in terms of median and $68\%$ confidence intervals (CI).

\section{Non-minimally coupled dark energy}
\label{sec:theory}

\subsection{Model and $\alpha_M$ parameter}

We consider a scalar field dark energy model described by the action
\begin{equation}\label{eq:action}
    S = \int d^4x\sqrt{-g}\left[\frac{M_{\rm Pl}^2}{2}F(\varphi)R
        + X - V(\varphi) + \mathcal{L}_M\right],
\end{equation}
where $M_{\rm Pl}^2$ is the Planck mass (and we will work in units where $M_{\rm Pl}^2=1$), $X=-\partial_\mu\varphi\partial^\mu\varphi/2$ is the familiar canonical kinetic term, and the non-minimal coupling function and potential function are \cite{Wolf:2025jed}
\begin{equation}
    F(\varphi)\simeq 1-\xi\varphi^2, \quad V(\varphi)\simeq
    V_0+\beta\varphi+ \frac{1}{2}m^2\varphi^2.
\end{equation}
The non-minimal coupling is quadratic in $\varphi$ while the form of the potential represents a Taylor expansion, which is motivated by the fact the field excursions within scalar field dark energy models are in most cases constrained to be small due to the scalar field being frozen by Hubble friction for significant portions of cosmic history and constraints that still favor equation of state values that have not evolved too far away from the $\Lambda$ value. Thus, this model is very general, and is completely degenerate with a huge variety of other potentials and dark energy models in terms of its resulting expansion history \cite{Wolf:2025jed, Wolf:2024stt, CGG_PGF_WJW_2026}. While minimally coupled quintessence ($\xi = 0$) is the most widely explored realization of single scalar field dark energy, the non-minimal coupling $\xi$ can be seen as naturally arising in the EFT of the scalar field \cite{Park:2010cw}. This coupling modifies the strength of gravity on cosmological scales and introduces notably different cosmological dynamics from that of its minimally coupled cousin \cite{Wolf:2025jed, Wolf:2024stt, Boisseau:2000pr, Gannouji:2006jm, Perrotta:1999am, Ye:2024ywg}:
\begin{equation}
    \ddot{\varphi}+3H\dot{\varphi}
+V'(\varphi)
-\frac{1}{2}F'(\varphi)R
=0.
\end{equation}
In the very early universe during radiation domination, $R\simeq 0$, the scalar field is frozen due to Hubble friction, and it behaves as in standard quintessence ($w_{\varphi}\simeq-1$). However, as the field begins to thaw, the pressure $p_\varphi
=
X
-V(\varphi)
-2\xi\dot{\varphi}^{2}
-2\xi\varphi\ddot{\varphi}
-4\xi H\varphi\dot{\varphi}$ and energy density $\rho_\varphi
=
X
+V(\varphi)
+6\xi H\varphi\dot{\varphi}$
of the field depart from the usual behavior in quintessence\footnote{There are different conventions for representing the additional dependence on $\xi$ and $\varphi$ that the effective pressure and energy density of the dark energy field possess, as well as for defining an effective equation of state in the presence of non-minimal couplings and modified gravity more generally. See e.g.~\cite{Bellini:2014fua, Madsen:1988ph, Gannouji:2006jm, Boisseau:2000pr} for further discussion.} due to the presence of the non-minimal coupling (which can be thought of as a curvature dependent mass with $m^2_{\rm eff} = m^2+\xi R$). This results in an effective equation of state that dips into the phantom regime and then crosses the phantom divide from below around $z\simeq0.5$. This behavior is indicative of a dark energy density that was increasing with cosmic expansion up to $z\simeq 0.5$, and then begins to decrease as it thaws during the most recent redshifts. Preference for dark energy with this behavior is seen in the cosmological data across a multitude of distinct analysis pipelines, including both parametric and model agnostic interpretations of the data \cite{DESI:2025fii, Berti:2025phi, Wolf:2025jlc}. 

The NMC model in Eq.\ (\ref{eq:action}) is a physically motivated, microphysical realization of dark energy that can naturally explain cosmological data. In particular, as shown in \cite{Wolf:2025jed, Wolf:2024stt, CGG_PGF_WJW_2026} when studying the combination of baryon acoustic oscillations (BAO) + cosmic microwave background (CMB) + supernova (SNe) data, the NMC model compares very favorably to the parametrized $w_0-w_a$ models and substantially outperforms the $\Lambda$CDM model in its ability to describe the cosmological data. Statistically, the model offers $\Delta\chi^2 \simeq -14 (\simeq3\sigma)$ to $-24 (\simeq 4.2\sigma)$, an Akaike Information Criterion (AIC) value of $\Delta \text{AIC} \simeq -8$ to -18, and a Bayes factor $\log B \simeq 3-7$ improvements over $\Lambda$, with the significance of the improvement depending on the exact supernovae set used in the analysis. However, recent recalibrations on the Union and DES SNe data \cite{DES:2025sig, Hoyt:2026fve} have settled around the $3.2\sigma-3.4\sigma$ range in terms of their preference for dynamical dark energy (when combined with CMB and BAO data), which end up on the more moderate side of these statistical measures. In this work, we make use of the latest constraints on the NMC model \cite{CGG_PGF_WJW_2026, Wolf:2025jed} that have been obtained from the combination of the DESI-DR2 BAO \cite{DESI:2025fii}, Planck PR3 CMB temperature and polarization power spectra \cite{Planck:2018vyg}, ACT+Planck lensing \cite{ACT:2023kun}, and 
DES-Dovekie \cite{DES:2025sig} SNe likelihoods in order to determine cosmological constraints on the parameter $\alpha_M$ and assess their compatibility with GW data.\footnote{We use the {\tt bao.desi\_dr2}, {\tt sn.desdovekie}, {\tt plik}, and {\tt actplanck\_baseline} likelihoods as implemented in {\tt cobaya} \cite{Cobaya, Cobaya2}.}

Before proceeding to this analysis, a word of caution is in order. Any theory that modifies gravity through a direct coupling of a field to gravity will introduce severe ancillary gravitational consequences. For example, a naive application of the theory Eq.~\eqref{eq:action} would lead to severe violations of solar system constraints on fifth forces and the time variation of the gravitational ``constant'' $G$ \cite{Uzan:2024ded}. However, Eq.~\eqref{eq:action} is an effective theory valid on \textit{cosmological} scales which are approximately $10^{13}-10^{15}$ larger than the solar system scales on which these constraints are valid. While we do not address this question here, there are a variety of plausible screening mechanisms that can shield dense environments like galaxies or the solar system from cosmological modified gravity \cite{Joyce:2014kja, Brax:2021wcv}. In particular, models with a conformal coupling as in  Eq.~\eqref{eq:action} are likely to be equipped with a chameleon screening mechanism, which would suppress both the radial and temporal modifications of the gravitational force in the Solar System \cite{Lagos:2020mzy}. 
As argued in \cite{CGG_PGF_WJW_2026}, we will need high quality cosmological data (e.g., related to the growth of structure, weak lensing, Integrated Sachs-Wolfe effect, etc) to unambiguously pin down gravity on cosmological scales. In this spirit, data constraining the  GW luminosity distance $D_L^{GW}$ offers us a novel and unique window into a very specific class of modified gravity that are characterized by their direct sensitivity to the cosmological evolution of the effective Planck mass $M^2_*$ \cite{Lagos:2019kds, Lagos:2020mzy, Wolf:2019hun, Belgacem:2018lbp, LISACosmologyWorkingGroup:2019mwx}.

In the scalar-tensor EFTDE language, dark energy perturbations are characterized by four redshift-dependent functions $\{\alpha_T, \alpha_K,\alpha_B,\alpha_M\}$ \cite{Bellini:2014fua}. The NMC model in Eq.\ (\ref{eq:action}) predicts
$\alpha_T=0$, $\alpha_K\neq 0$, $\alpha_B\neq 0$, $\alpha_M\neq 0$, with the effective Planck mass running parameter $\alpha_M$ given by\footnote{More generally we have that, \begin{equation}
M_*^2(\varphi, X, H) \equiv 2\left(G_4-2 X G_{4 X}+X G_{5 \varphi}-\dot{\phi} H X G_{5 X}\right),
\end{equation}
where $G_4$ and $G_5$ are the Horndeski functions \cite{Bellini:2014fua}. Measurements of the luminous speed of gravitational waves \cite{LIGOScientific:2017vwq} have put incredibly tight constraints on $G_{4X}$, $G_{5\varphi}$, and $G_{5X}$ \cite{Baker:2017hug, Ezquiaga:2017ekz}, leaving $G_4$ as the only remaining parameter with substantial freedom. In our model $G_4(\varphi)=F(\varphi)/2$.}:
\begin{equation}
    \alpha_M(z) \equiv \frac{d\ln M_*^2}{d\ln a}
    = \frac{-2\xi\varphi\dot\varphi/H}{1-\xi\varphi^2}.
\end{equation}
For the best-fit parameters of Wolf et al.\ \cite{Wolf:2025jed}, $\alpha_M(z)$ is negative at low
redshift and approaches zero at high $z$, reflecting a decrease in the
effective Planck mass as the Universe evolves. 

\begin{figure}[h!]
	\centering
	\includegraphics[width=\columnwidth]{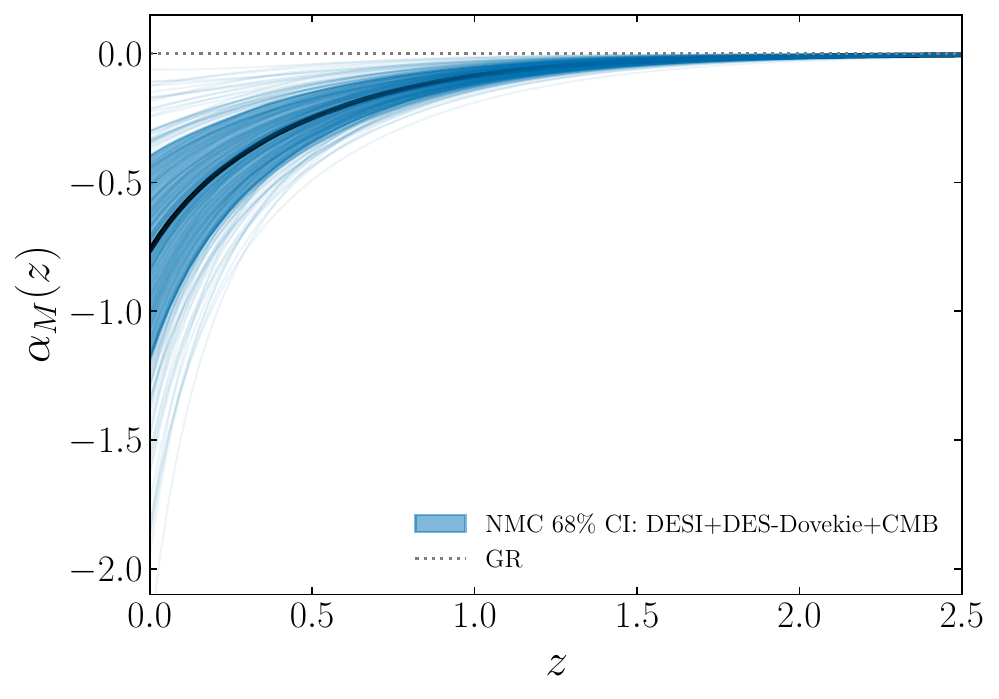}
	\caption{Posterior distribution of $\alpha_M(z)$ from the NMC
	quintessence model of Wolf et al.~\cite{Wolf:2025jed} (selection out of 2500 samples,
	blue traces) determined from the combination of DESI-DR2 BAO \cite{DESI:2025fii}, Planck PR3 CMB temperature and polarization power spectra \cite{Planck:2018vyg},  ACT+Planck lensing \cite{ACT:2023kun}, and DES-Dovekie SNe \cite{DES:2025sig}. The black solid line is the posterior median; the blue
	band is the 68$\%$ CI. The grey dotted line marks General Relativity
	($\alpha_M=0$). }
	\label{fig:alphaM}
\end{figure}

Fig.~\ref{fig:alphaM} shows the posterior of $\alpha_M(z)$ as constrained by BAO+SNe+CMB cosmological data. The function is consistently negative at low redshift, peaks in
magnitude at $z=0$, and decays to zero by $z\sim 2$.  This behavior
reflects the decreasing effective Planck mass driven by the non-minimal coupling
$\xi\varphi^2 R$ with $\alpha_M(z=0)=-0.76^{+0.37}_{-0.42}$ today (median and $68\%$CI), confirming a statistically significant preference for negative $\alpha_M$ at low redshift. In particular, the $\Lambda$CDM value $\alpha_M(z=0)=0$ lies outside the $99.5\%$CI. 

While $\alpha_M$ is the most relevant parameter for our purposes as it is this term that impacts the cosmological propagation of gravitational waves, we also note that for the NMC model we have that $\alpha_M=-\alpha_B$ (as can be easily derived from the $\alpha$ expressions in \cite{Bellini:2014fua}) and that $\alpha_K$ is generally a small positive value. However, $\alpha_K$ has only a minimal impact on cosmological observables and is not further considered here, while $\alpha_B$ is important for many effects relevant to modified gravity and will play a role in confronting the EFTDE parameters to cosmological data later.

\subsection{GW propagation parametrizations}

\begin{figure*}
\centering
\includegraphics[width=0.49\textwidth]{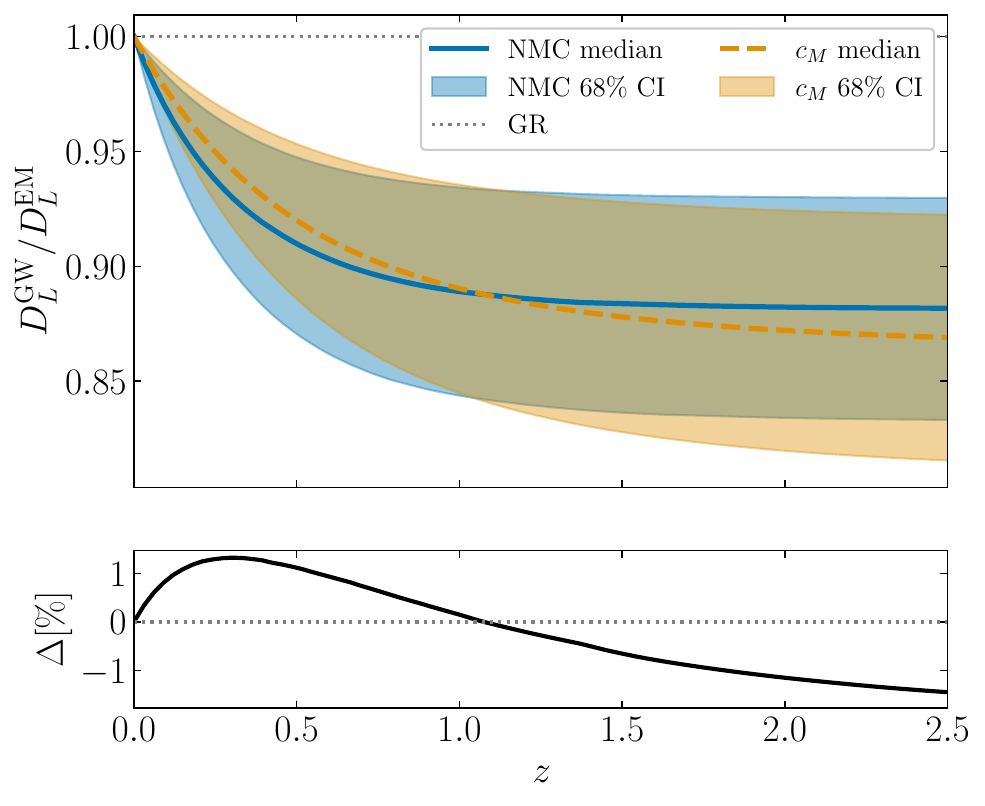}
\includegraphics[width=0.49\textwidth]{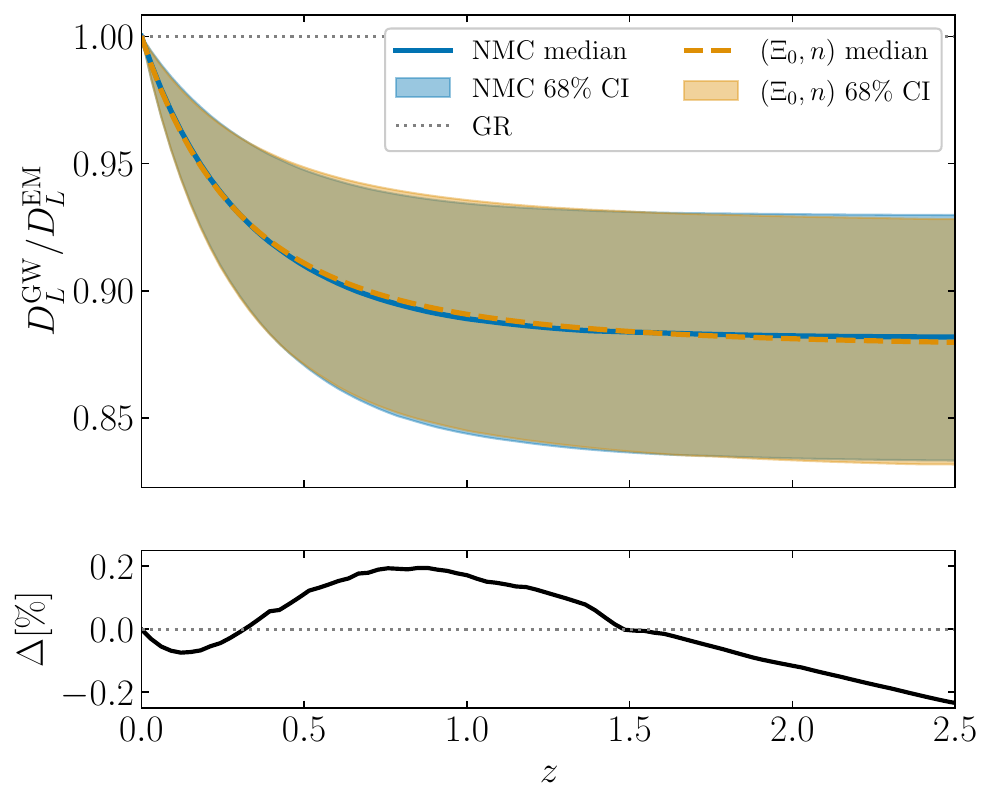}
\caption{Left: $D_L^{\rm GW}/D_L^{\rm EM}$ vs.\ redshift predicted by the NMC model and its $c_M$ parametrization fit.  Blue traces and band: NMC median and $68\%$CI of the posterior. Orange dashed line and band: median and $68\%$CI from the $c_M$ model fit. Right: Same as the left panel but for the $(\Xi_0,n)$ parametrization fit. In both right and left panels, we also show the residuals between the two curves. Black dotted line is the GR prediction. }
\label{fig:ratio}
\end{figure*}

When studying the cosmological propagation of gravitational waves in conformal time, the equation of motion of a monochromatic wave with wavenumber $k$ is given by:
\begin{equation}
    h_A^{''}+[2+\alpha_M(z)]\mathcal{H}h_A'+k^2h_A=0,
\end{equation}
where $A$ is a subscript describing the two GW polarizations $+,\times$, $\mathcal{H}$ is the conformal Hubble factor, and $\alpha_M$ is a generally redshift-dependent modification to the $\Lambda$CDM prediction. This modification induces changes in the amplitude decay of GWs when propagating from the source to the observer. In the $\Lambda$CDM model ($\alpha_M=0$) the GW amplitude decays with the standard luminosity distance $D_L^{EM}$ as $h_A\propto 1/D_L^{\rm EM}$, but if $\alpha_M\not=0$ then GWs will decay with a new prefactor called the `GW luminosity distance' $h_A\propto 1/D_L^{\rm GW}$.

In the literature, it is customary to parametrize either $\alpha_M$ or $D_L^{\rm GW}$ in order to perform theory-agnostic tests of cosmology. Two parametrizations are commonly used in GW analyses \cite{LIGOScientific:2026uyd}.

\paragraph{The $c_M$ parametrization \cite{Lagos:2019kds}.}
This is a 1-parameter model that assumes $\alpha_M$ to evolve with the fractional dark energy density as in the $\Lambda$CDM  model:
\begin{equation}
    \alpha_M(z) = c_M\frac{\Omega_\Lambda(z)}{\Omega_{\Lambda0}}
    = \frac{c_M}{E^2(z)},
    \label{eq:cM}
\end{equation}
where 
\begin{equation}
    E^2(z)=\Omega_m(1+z)^3+\Omega_{\Lambda0}.
\end{equation}
Here, $\Omega_{m,\Lambda0}$ are the fractional energy densities of non-relativistic matter and dark energy today, and $c_M$ is a constant free parameter, with $c_M=0$ recovering General Relativity (GR) as assumed in the standard $\Lambda$CDM cosmological model. We use the
flat-$\Lambda$CDM relation $\Omega_{\Lambda 0}=1-\Omega_m$.

 In this case, the derived GW distance ratio is \cite{Lagos:2019kds, Belgacem:2018lbp}
\begin{equation}
	\frac{D_L^{\rm GW}}{D_L^{\rm EM}}
	= \exp\!\left(\frac{1}{2}\int_0^z\frac{dz'}{1+z'}\,\alpha_M(z')\right).
	\label{eq:ratio}
\end{equation}

The left panel of Fig.~\ref{fig:ratio} shows the resulting $D_L^{\rm GW}/D_L^{\rm EM}$
ratio obtained in the NMC model by integrating each $\alpha_M^{(i)}(z)$ curve through
Eq.~(\ref{eq:ratio}), and a comparison with the $c_M$ parametrization fit.  The ratio drops steeply from unity at low $z$ and approaches an asymptotic value of $\simeq 0.88$ by $z\sim 1.5$.  Since
$\alpha_M<0$, GW sources appear systematically \emph{closer} than their
EM counterparts would predict. 

We fit the $c_M$ parametrization to the NMC $D_L^{\rm GW}/D_L^{\rm EM}$ posterior minimizing a $\chi^2$ statistic discussed in Appendix \ref{app:fitting}, which is an ordinary least squares function reweighted by the fractional error of the NMC data at each redshift. The fit yields
\begin{equation}\label{cM_value}
	c_M = -0.5\pm 0.2 \quad (68\%\ \mathrm{CI}),
\end{equation}
which is in approximately $2.4\sigma$ tension with GR.
The negative sign follows directly from the NMC having $\alpha_M<0$.
The left panel of Fig.~\ref{fig:ratio} shows that the $c_M$ model captures the overall shape
of the ratio, though it falls slightly above the data median at
low redshifts ($z\lesssim 1$) and then below for higher redshifts ($z\gtrsim 1$), reflecting the
mismatch between the $c_M/E^2$ template and the more complex
$z$-dependence of the NMC $\alpha_M$ curve. Note that this parametrization assumes that $\Omega_\Lambda(z)$ behaves as in the $\Lambda$CDM model, whereas the NMC model predicts a more complex behavior. 

In the bottom left panel of Fig.~\ref{fig:ratio} we show the fractional deviation $\Delta$ between the median distance ratio of the NMC model and that of the $c_M$ fitted parametrization. We can see that the deviations can be at the percent level.  

\paragraph{The $(\Xi_0,n)$ parametrization \cite{Belgacem:2018lbp}.}
In this case, the distance ratio is parametrized directly as: 
\begin{equation}
    \frac{D_L^{\rm GW}}{D_L^{\rm EM}}
    = \Xi_0 + \frac{1-\Xi_0}{(1+z)^n},
    \label{eq:Xi0n}
\end{equation}
which has two positive free parameters $\Xi_0>0$ and $n>0$. The GR limit $\Xi_0=1$ (for any $n$). This parametrization is constructed such that at low redshift $D_L^{\rm GW}/D_L^{\rm EM}\rightarrow 1$ and at high redshift it asymptotes to $D_L^{\rm GW}/D_L^{\rm EM}\rightarrow \Xi_0$ since the modifications should saturate at redshifts where dark energy becomes irrelevant. In this model, the power index $n$ controls the redshift evolution and, as we will see next, this will allow to fit better the NMC model compared to the $c_M$ parametrization that has a fixed redshift evolution.

The right panel of Fig.~\ref{fig:ratio} shows the resulting $D_L^{\rm GW}/D_L^{\rm EM}$ fit obtained.
Fitting the $(\Xi_0,n)$ model gives
\begin{equation}\label{xi0n_values}
	\Xi_0 = 0.88\pm 0.05,\quad
	n = 3.2\pm 0.3 \quad (68\%\ \mathrm{CI}).
\end{equation}
GR ($\Xi_0=1$) is disfavored at approximately $ 2.5\sigma$. We notice that the residuals are ten times lower than those from the $c_M$ parametrization, confirming that this 2-parameter $(\Xi_0,n)$ model can fit better the complex $z$-dependence predicted by the NMC model.

\begin{figure}[h!]
	\centering
	\includegraphics[width=\columnwidth]{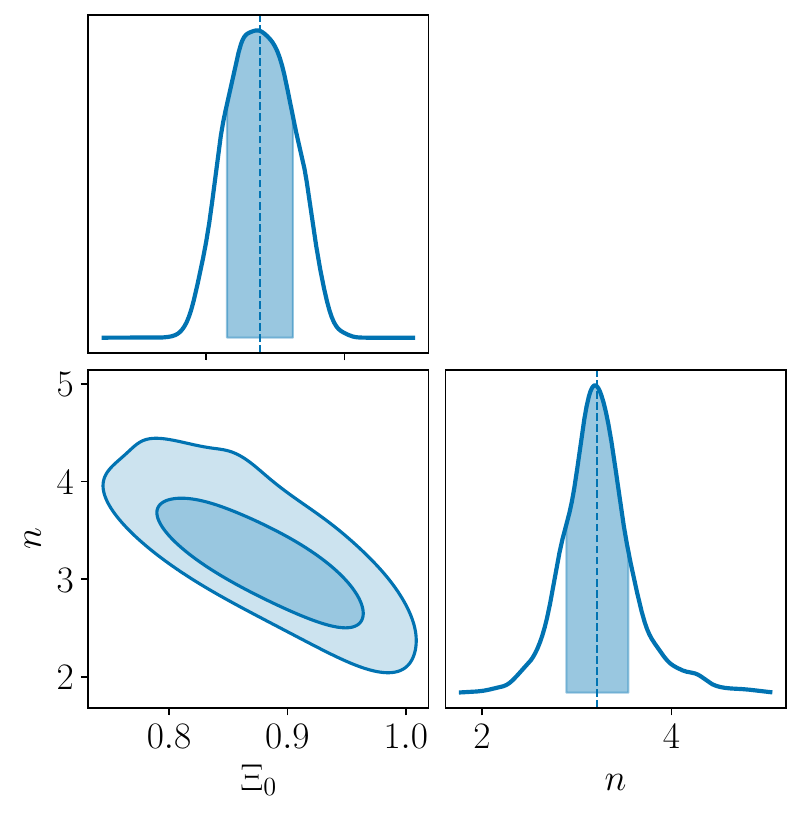}
	\caption{Joint posterior $p(\Xi_0,n)$ from fitting the
		$(\Xi_0,n)$ parametrization to the NMC model. Marginalized posteriors show $68\%$ CI shaded, and joint posterior shows both $68\%$ and $95\%$ CI contours. }
	\label{fig:corner}
\end{figure}

The joint posterior
(Fig.~\ref{fig:corner}) shows a clear negative correlation (Pearson correlation coefficient $r=-0.7$): larger $\Xi_0$ (closer to GR) requires a shallower $z$-dependence (lower $n$) to reproduce
the low-$z$ drop in the distance ratio.

\paragraph{Parametrization Comparison.}
In order to assess how well the $c_M$ and $(\Xi_0,n)$ parametrizations fit the NMC model, we define a $\chi^2$ goodness of fit statistics per each sample line in  Fig.\ \ref{fig:alphaM}, given by:
\begin{equation}\label{eq:chi2}
 \chi_i^2(\vec{p})=\sum_j \frac{\left(R_{i}(z_j)-R_{\rm model}(z_j;\vec{p})\right)^2}{\sigma_r(z_j)^2}   .
\end{equation}
In order to propagate the full NMC posterior into the parametrized description, we fit each posterior sample individually rather than a single summary curve. For each sample line $i$ in Fig.\ \ref{fig:alphaM}---whose associated distance ratio is $R_i=D_L^{GW}/D_L^{EM}$---we obtain the best-fit parameters $\hat{\vec{p}}$ by minimizing $\chi^2_i$ in Eq.\ (\ref{eq:chi2}). Here, we add the NMC model samples at each redshift approximated by independent gaussian distributions, where $\sigma_r$ corresponds to the fractional $68\%$ uncertainty of the samples at redshift $z_j$ \footnote{\label{foot:sigma}Because the per-redshift NMC posteriors of $R$ are not exactly gaussian, the 16th/84th percentiles may yield different upper and lower $68\%$ uncertainties, $\sigma^{+}(z_j)$ and
$\sigma^{-}(z_j)$. In Eq.\ (\ref{eq:chi2}) we adopt the directional error,
using $\sigma^{+}$ when the sampled point lies above the posterior
median and $\sigma^{-}$ when it lies below. These errors are then normalized by the median of the samples at a given redshift $\bar{R}_{\rm NMC}(z_j)$ to obtain the fractional uncertainty $\sigma_r$.}. Finally, the model template $R_{\rm model}(z_j;\vec{p})$ will depend on the model parameters $\vec{p}$ (either $c_M$ or $(\Xi_0,n)$).
More details can be found in Appendix \ref{app:fitting}. We note that in Eq.\ (\ref{eq:chi2}) we use the fractional error as normalization to give more weight to redshift regions that have more constraining cosmological data, although we find $\sigma_r$ to be quite constant between the redshift region we consider $0<z<2.5$.

Based on this statistic, we obtain a distribution for $\chi^2$, whose median value is: $\chi^2=4.2$ for the $c_M$ parametrization and $\chi^2=0.04$ for the $(\Xi_0,n)$ parametrization. These numbers confirm the visual result in Fig.\ \ref{fig:ratio} that the  $(\Xi_0,n)$ model adjusts the NMC data much better than the $c_M$ model.

\section{GW Cross-Probe Consistency}
\label{sec:GW_agreement}

A central question in contemporary cosmology is whether the growing evidence for dynamical dark energy points toward a specific microphysical model, and whether such a model produces consistent signatures across independent observational probes. The NMC model of Wolf et al. \cite{Wolf:2025jed} is particularly well-suited to address this question, as its non-minimal coupling to gravity generates not only a modified expansion history---already shown to be strongly preferred by CMB, BAO, and SNe data---but also a distinct and unavoidable signature in gravitational wave propagation through the GW friction function $\alpha_M(z)$. This means that the same model that fits the background cosmological data makes concrete, falsifiable predictions for the GW luminosity distance ratio $D^{\rm GW}_L/D^{\rm EM}_L$, which can be tested independently with standard sirens.

This cross-probe consistency test is non-trivial because GW observations and CMB+LSS+SNe observations are subject to entirely different systematic effects, so agreement between them would provide evidence that a time-varying dark energy model is genuine rather than a statistical fluctuation or systematic artifact. In this section, we compare the NMC model predictions for $c_M$ and $(\Xi_0,n)$ against the GWTC-5 GW constraints \cite{LIGOScientific:2026uyd}.

\subsection{GW Data}

The GWTC-5 catalog \cite{LIGOScientific:2026uyd} contains 236 gravitational-wave sources detected by the LIGO–Virgo–KAGRA collaboration across observing runs O1 through O4b, selected with a false alarm rate below 0.25 per year. For the modified GW propagation constraints relevant to this work, GWTC-5 relies exclusively on the spectral siren method\footnote{The LVK collaboration found the dark-siren modified-gravity results insufficiently robust (see footnote 3 in  \cite{LIGOScientific:2026uyd}).}, where redshift information is extracted from features in the source-frame mass distribution of 235 compact binary coalescences (excluding the only bright siren event GW170817---which, on its own, yields large uncertainties on the running Planck mass \cite{Lagos:2019kds} due to its low redshift).

GWTC-5 constrains parameterized deviations from GR affecting GW propagation, using both the $c_M$ and $(\Xi_0,n)$ parametrizations described in the previous section. These constraints, summarized in Table 4 of \cite{LIGOScientific:2026uyd}, provide the GW-side comparison point for our analysis. Since $H_0$ and the modified GW propagation parameters can be correlated---both affecting the luminosity distance–redshift relation---the GWTC-5 analysis presents constraints under two different priors. A wide prior, $H_0\in [10,120]$ km/s/Mpc, takes the most agnostic approach and allows the GW data alone to determine the degeneracy between $H_0$ and the propagation parameters. A narrow prior, $H_0\in [65,77]$ km/s/Mpc, instead assumes prior knowledge of the Hubble constant consistent with current CMB and local distance-ladder measurements.

\subsection{Agreement Analysis: NMC vs.\ GWTC-5}
\label{subsec:gw_tension}
To quantify the agreement between two posterior distributions $1$ and $2$
we use: 
\begin{equation}
\mathcal{T}=\frac{|\mathrm{med}_1-\mathrm{med}_2|}{\sqrt{\sigma^2_{1\rightarrow 2}+\sigma^2_{2\rightarrow 1}}},\label{eq:T}
\end{equation}
where we are taking the difference of the median of two distributions, and normalize them by their $68\%$ uncertainties (16th-84th percentiles), taken directionally toward the other distribution's median. This generalizes the usual Gaussian tension estimator to asymmetric posteriors: because the uncertainties are defined from the 16th/84th percentiles rather than a symmetric standard deviation, they follow the asymmetry of each distribution. Notice that for the purposes of interpreting this one-parameter tension statistic, we can directly interpret its value as an approximate $n-\sigma$ tension.

We compute this statistic $\mathcal{T}$ after normalizing both
distributions via kernel density estimation (KDE) for our posterior samples. For GWTC-5, we use publicly available data on $c_M$ and joint $(\Xi_0,n)$ posteriors \cite{LVK_GWTC5_2026}.

\begin{figure}[h!]
    \centering
    \includegraphics[width=\columnwidth]{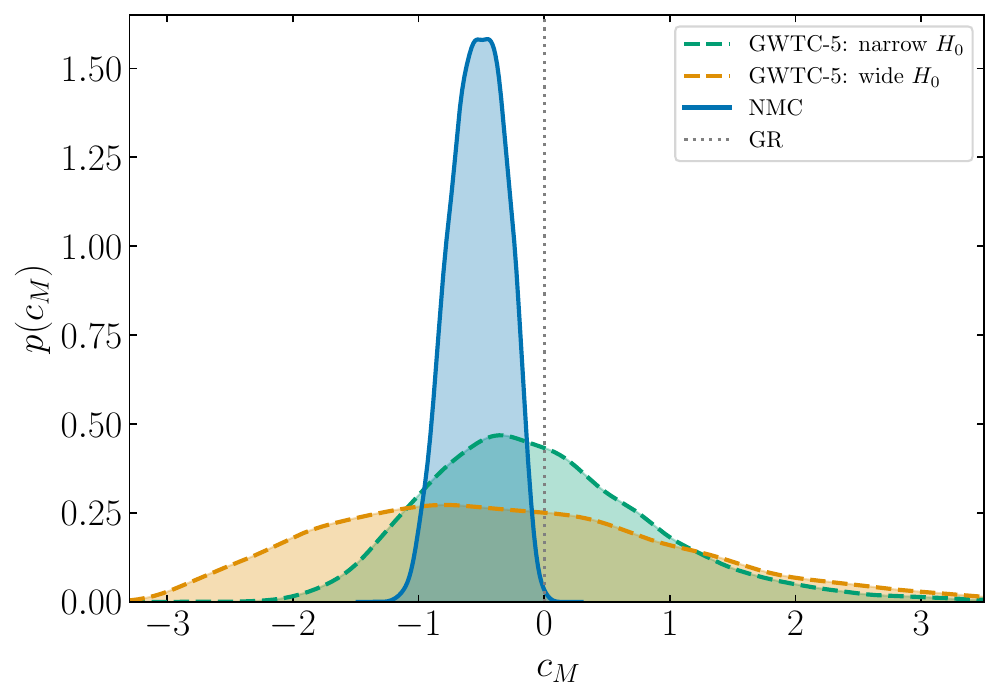}
    \caption{Comparison of $c_M$ posteriors between the NMC model and GW data. \textit{Blue solid}:
    inferred from the NMC posterior (this work).
    \textit{Green dashed}: GWTC-5 GW constraint for
    narrow $H_0$ prior.
    \textit{Orange dashed}: GWTC-5 GW constraint for
    wide $H_0$ prior.  The vertical
    grey dot-dashed line marks GR. }
    \label{fig:GW-cM_compare}
\end{figure}

Fig.\ \ref{fig:GW-cM_compare} shows the $c_M$ posterior, comparing the NMC model to the GW standard siren data for narrow and wide $H_0$ priors. The GWTC-5 constraints for narrow and wide $H_0$ priors are  $c_M=-0.1^{+1.0}_{-0.8}$ and $c_{M}=-0.4^{+1.6}_{-1.3}$, respectively.

Here we see that that these GW constraints are much wider than the one obtained from cosmological data for the NMC model. By using Eq.\ (\ref{eq:T}) we obtain $\mathcal{T}=0.5$ and $\mathcal{T}=0.08$ for NMC-GWTC-5 narrow and wide, respectively. Here, the narrow case gives a larger tension because its median is further from the NMC median and the uncertainties are smaller. Nonetheless, a value of $\mathcal{T}=0.5$ shows great consistency between the NMC model and GW data.

\begin{figure}[h!]
    \centering
    \includegraphics[width=\columnwidth]{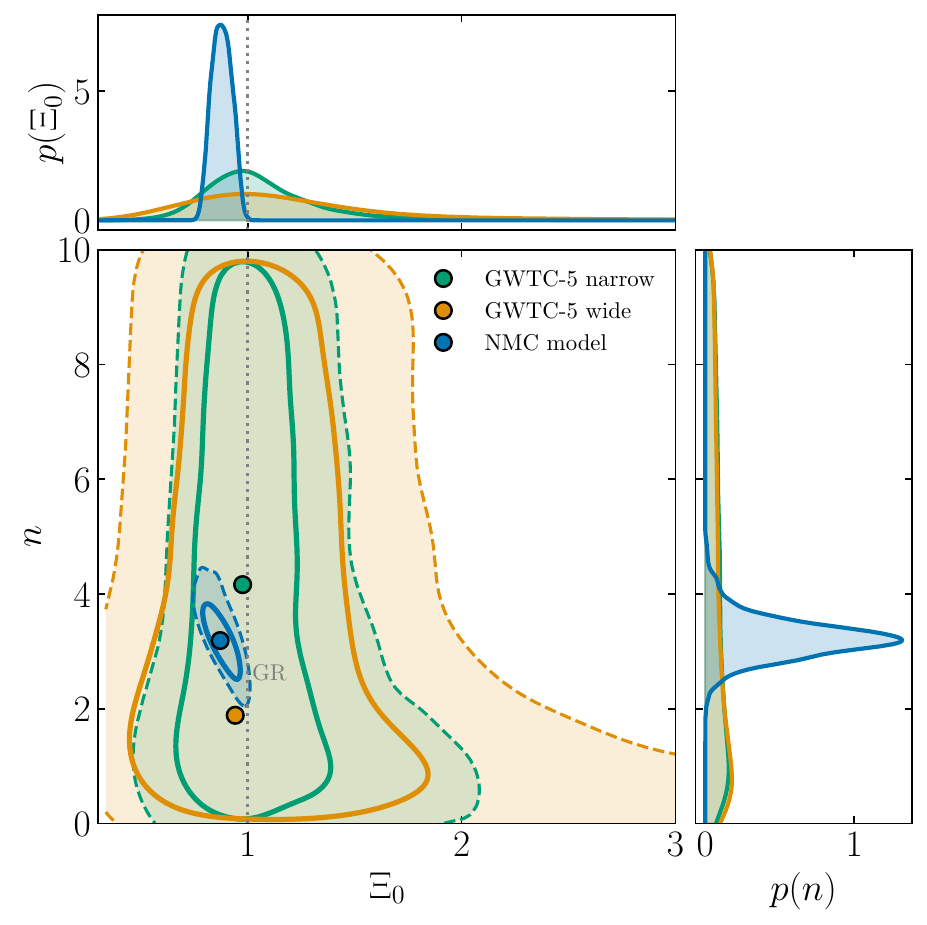}
    \caption{Comparison of $(\Xi_0,n)$ posteriors between the NMC model and GW data. \textit{Blue}:
    inferred from the NMC  posterior (this work).
    \textit{Green}: GWTC-5 GW constraint for
    narrow $H_0$ prior.
    \textit{Orange}: GWTC-5 GW constraint for
    wide $H_0$ prior. Solid and dashed contours describe $68\%$ and $95\%$ CI, respectively. The vertical grey dot-dashed line marks GR. }
    \label{fig:GW-Xi0_compare}
\end{figure}

Fig.\ \ref{fig:GW-Xi0_compare} shows the $(\Xi_0,n)$ posteriors, comparing the NMC model to the GW standard siren data for narrow and wide $H_0$ priors. The GWTC-5 constraints for narrow and wide $H_0$ priors are $(\Xi_0,n)=(1.0^{+0.3}_{-0.2}, 3.8^{+3.8}_{-2.8})$ and $(\Xi_0,n)=(1.1^{+0.6}_{-0.3}, 3.4^{+4.2}_{-2.6})$, respectively.

Since the GWTC-5 posteriors on $n$ are quite wide, we measure $\mathcal{T}$ only through the marginalized $\Xi_0$ distribution. In this case, we obtain  $\mathcal{T}=0.7$ and $\mathcal{T}=0.7$ for narrow and wide $H_0$ priors, respectively, which shows agreement between the two data sets.

Overall, the NMC predictions for $c_M$ and $(\Xi_0,n)$ are consistent with the current GWTC-5 GW constraints. However, this agreement should not be over-interpreted: it is largely a consequence of the fact that current GW constraints remain very wide, with a $1\sigma$ uncertainties these parameters of order unity, comfortably encompassing the NMC predictions but also the predictions of GR. In other words, the GW data are not yet precise enough to distinguish the NMC model from GR, let alone from other dark energy models. The consistency we find is therefore an encouraging but not yet stringent cross-probe validation of the NMC model.

GW forecasts show that 100 binary neutron star bright sirens detected by LIGO at design sensitivity would reach $\sigma(c_M)\approx 0.9$ when assuming $H_0$ is not known or  $\sigma(c_M)\approx 0.5$ when assuming $H_0$ is known \cite{Lagos:2019kds, Mastrogiovanni:2020gua}. These estimates show that 100 events would still not be enough to confidently distinguish GR from the NMC model ($c_M\approx -0.5$), and one would need  approximately $ 10^3$ bright siren events in order to have $\sigma (c_M)=0.1$ and distinguish GR from the NMC model with $5\sigma$ confidence, when assuming $H_0$ is known. Regarding dark sirens studies, \cite{Leyde:2022orh} shows that 510 BBH events detected in by LVK with O4 and O5 sensitivity, would reach $\sigma (\Xi_0)\approx 0.4$ while $n$ would remain largely unconstrained, when agnostic about $H_0$. Such results are comparable to current GWTC-5 constraints by LVK \cite{LIGOScientific:2026uyd}, which we have shown to be unable to distinguish between the NMC model and GR. Instead, with nearly 5000 BBH events observed at design sensitivity, constraints could reach a precision of $\sigma (\Xi_0)\approx 0.14$ \cite{Mancarella:2021ecn}, which would still not be enough to distinguish GR from the NMC model.

Next-generation GW detectors such as Einstein Telescope (ET) and Cosmic Explorer (CE) are expected to dramatically improve this situation: \cite{Belgacem:2018lbp} estimates that $10^3$ bright sirens detected by ET could lead to measurements of $\Xi_0$ with $1\%$ precision, at which point the NMC value $\Xi_0\approx 0.88$ could be distinguishable from GR and assessed at high significance. 
In the case of the space-based detector LISA, forecasts on bright sirens estimate that $\Xi_0$ can be measured to percent-level precision \cite{LISACosmologyWorkingGroup:2019mwx} as well\footnote{The percent-level forecast on $\Xi_0$ from \cite{LISACosmologyWorkingGroup:2019mwx} is obtained with $n$ fixed to values motivated by specific modified gravity models, rather than marginalized over. The NMC model predicts $n \approx 3.2$, which differs from the values assumed in those forecasts. A new dedicated forecast  would be needed to consistently assess the precision with which LISA could constrain $\Xi_0$ in this specific NMC model.}.

Should future high-precision GW measurements yield results in tension with the NMC predictions, a careful assessment of the differences between the analyses would be required before drawing firm conclusions. In particular, the GWTC-5 modified gravity constraints assume a flat-$\Lambda$CDM expansion history: with a constant $\Omega_\Lambda$ and $\Omega_m=0.3065$ fixed. However, the NMC model predicts a modified background dynamics, with an evolving fractional dark energy density $\Omega_\phi(z)$ and a best-fit value for $\Omega_m=0.311$.
In principle, a fully consistent comparison would therefore require a joint parameter estimation analysis where cosmological+GW data is simultaneously fitted for the expansion history and the GW friction function, in the context of the NMC model. We leave such a dedicated joint analysis for a future work.

\subsection{$H_0$ Bias}\label{sec:H0bias}

As found in \cite{Lagos:2019kds}, a degeneracy between $H_0$ and $c_M$ is present in the analysis of standard sirens. 
If the NMC model is correct with $c_M\simeq -0.5$
but GW analyses assume $c_M=0$ (GR), the inferred GW distances are biased, and the resulting $H_0$ value is correspondingly shifted.

\begin{figure}[h!]
	\centering
	\includegraphics[width=\columnwidth]{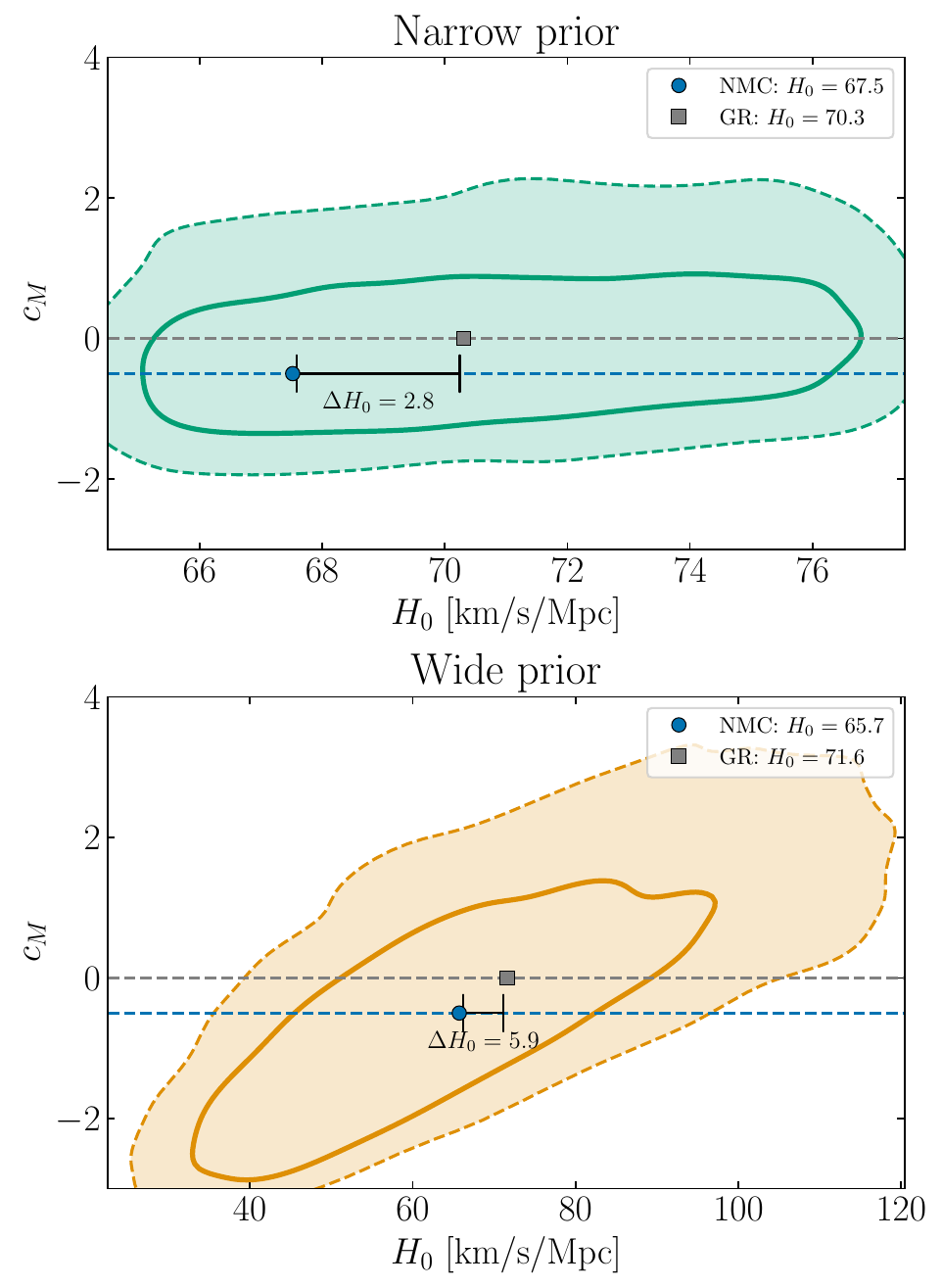}
	\caption{Joint posterior distribution of $H_0-c_M$ according to GWTC-5 data, assuming narrow (top) and wide (bottom) $H_0$ priors. Dots and squares show the best-fit value of $H_0$ when $c_M$ is -0.5 or 0, respectively. }
	\label{fig:H0-cMgw}
\end{figure}

In Fig.\ \ref{fig:H0-cMgw} we show the GWTC-5 joint $H_0-c_M$ posterior from standard sirens, for narrow (top) and wide (bottom) $H_0$ priors. GWTC-5 data reveal a clear positive degeneracy between the two parameters: larger $H_0$ is preferred alongside larger $c_M$. We highlight the best-fit value for $H_0$ when the true $c_M$ value is either -0.5 (as predicted by the NMC model) or 0 (as predicted by GR). 

Fixing $c_M=-0.5$ and reading off the best-fit $H_0$ along the posterior ridge, we find $H_0 = 67.5$ km/s/Mpc under the narrow prior, compared to $H_0=70.3$ km/s/Mpc when GR is assumed ($c_M = 0$)---a shift of $\Delta H_0=2.8$ km/s/Mpc. Under the wide prior, where the degeneracy is less constrained by the prior boundary, and thus the shift is larger: $H_0 = 65.7$ km/s/Mpc at $c_M = -0.50$ versus $H_0 = 71.6$ km/s/Mpc at $c_M = 0$, corresponding to a shift of $\Delta H_0 = 5.9$ km/s/Mpc.

\begin{figure}[h!]
	\centering
	\includegraphics[width=\columnwidth]{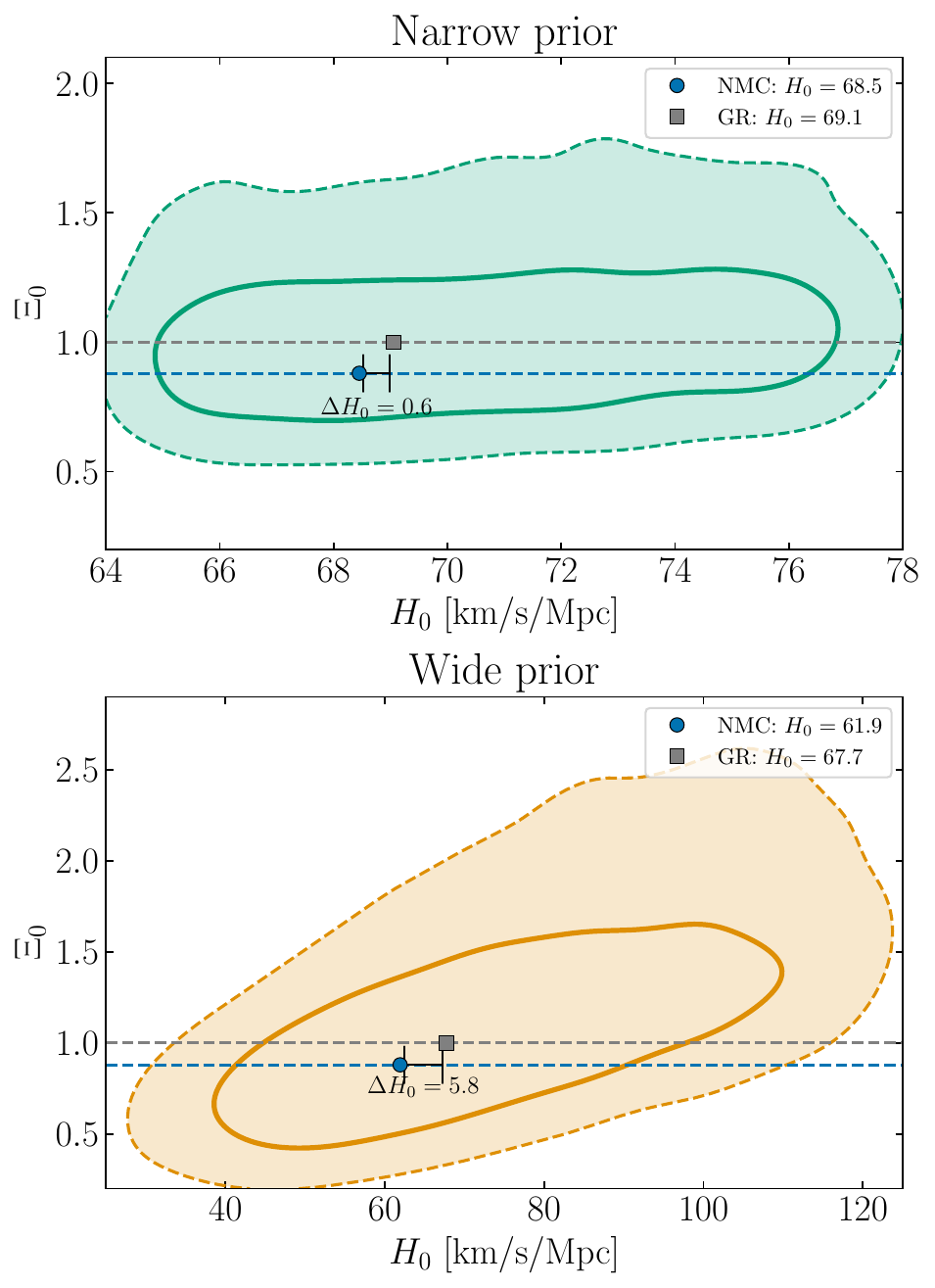}
	\caption{Joint posterior distribution of $H_0-\Xi_0$ according to GWTC-5 data, assuming narrow (top) and wide (bottom) $H_0$ priors. Dots and squares show the best-fit value of $H_0$ if $\Xi_0$ is 0.88 or 1, respectively.}
	\label{fig:H0-Xi0gw}
\end{figure}

Notably, the current Hubble tension \cite{DiValentino:2021izs} between the best-fit Planck 2018 value \cite{Planck:2018vyg}  ($67.8$ km/s/Mpc) and local distance-ladder measurements by SH0ES \cite{Riess:2021jrx} ($73$ km/s/Mpc) is comparable to the bias we estimate here. A GW analysis that properly accounts for the NMC $c_M$ would therefore shift $H_0$ downward relative to the GR assumption, moving it closer to Planck's value and thus alleviating the tension between early and late-time universe measurements.

In the case of the $(\Xi_0,n)$ parametrization, the correlation between $\Xi_0$ and $H_0$ in standard sirens analyses is somewhat weaker, as shown in Fig.\ \ref{fig:H0-Xi0gw}. In the case of a narrow prior, the best-fit $H_0$ value in the NMC model would be $68.5$ and it only shifts by $\Delta H_0=0.6$km/s/Mpc, giving no significant bias. For the wider prior there is a larger shift of $\Delta H_0=5.8$km/s/Mpc. Importantly, again $H_0$ is biased towards higher values when assuming GR.

\section{Cosmological Models Consistency}
\label{sec:Shah_agreement}
As highlighted by Shah et al.\ \cite{Shah:2025vnt}, the EFTDE parametrizations commonly used in both cosmological and GW analyses impose non-trivial theoretical priors on the space of dark energy models, meaning that constraints derived within these frameworks can be misleading when the true underlying dark energy model has a different structure. It is therefore important to assess not only whether the NMC model is consistent with current GW data, but also whether the standard parametrized constraints from cosmological observations are themselves a faithful representation of the NMC model's predictions. In this section we
compare the NMC predictions for $c_M$ and the EFTDE cosmological constraints of Shah et al.\ \cite{Shah:2025vnt}.

\subsection{Shah et al.\ Model and Data}

Shah et al.\ \cite{Shah:2025vnt} investigate the effect of different theoretical priors on observational constraints on dark energy and modified gravity, focusing in particular on the dynamics of linear perturbations around cosmological backgrounds. Following the EFTDE, their approach decouples the background expansion history from the perturbation dynamics, a separation that is consistent with large classes of scalar-tensor theories where sufficient freedom exists for these two sectors to be parametrized independently.

Among the tests they perform, they parametrize the background expansion through the CPL equation of state $w(a) = w_0 + w_a(1 -a)$, while the perturbations are described within the Effective Field Theory of Dark Energy (EFTDE) framework \cite{Bellini:2014fua} in terms of the braiding parameter $\alpha_B$ and the running Planck mass parameter $\alpha_M$. The tensor speed parameter $\alpha_T$ is fixed to zero motivated by the GW170817 constraint \cite{LIGOScientific:2017zic,LIGOScientific:2017ync}, and the kineticity parameter $\alpha_K$ fixed to a small fiducial value as it has negligible impact on sub-horizon observables. 
For the non-vanishing perturbation functions $\alpha_{B,M}$ they consider two temporal evolutions: $\alpha_{B,M}\propto \Omega_\Lambda(a)$ and $\alpha_{B,M}\propto a$. In this section we discuss the comparison of the former parametrization $\alpha_{B,M}\propto \Omega_\Lambda(a)$ with the NMC model. This is the most widely adopted time dependence in the literature (as considered in the standard sirens analysis in Sec.\ \ref{sec:GW_agreement}), and the one for which Shah et al.\ provide the NMC-relevant $\alpha_B=-\alpha_M$ constraint. This relationship arises from requiring the stability of dark energy perturbations on gravitational wave backgrounds sourced by loud binary black hole mergers \cite{Creminelli:2019kjy}, and it is relevant for the NMC model since one can easily show that the model in Eq.\ (\ref{eq:action}) predicts $\alpha_M = -\alpha_B$. Shah et al.\ also consider the $\alpha_{B,M}\propto a$ evolution but only for the unconstrained case (independent $\alpha_B$ and $\alpha_M$), so we do not analyze it here.

The datasets used in the Shah et al.\ \cite{Shah:2025vnt}  analysis are: SDSS DR16 joint BAO + f$\sigma_8$ growth measurements determined from redshift space distortions (RSD)  \cite{eBOSS:2020yzd}, Planck CMB temperature and polarization power spectra and Planck lensing \cite{Planck:2018vyg, Planck:2019nip},  PantheonPlus SNe \cite{Scolnic:2021amr}, and the Integrated Sachs-Wolfe (ISW) effect likelihood \cite{Seraille:2024beb}  constructed from cross-correlations of the Planck CMB temperature with galaxy number counts from several photometric surveys. Based on this data, they find $c_M=0.12\pm 0.09$\footnote{Note that Shah et al.\ report the mean and $68\%$CI as $c_M=0.12^{+0.08}_{-0.10}$. The slight difference with the value we report is that we directly used the samples to obtain the uncertainties, using \texttt{numpy.percentile}, whereas Shah et al.\ used the \texttt{GetDist} package which smooths the samples with KDE.}. We notice that a recent reanalysis with DESI data was done by Shah et al.\  \cite{Shah:2026etz} but only considering the unconstrained case where $\alpha_M$ and $\alpha_B$ are independent parameters, and hence we do not discuss those results further here.

\subsection{Agreement Analysis: NMC vs.\ Shah et al.}
In this section, we perform an analogous agreement analysis to the one in Sec.\ \ref{subsec:gw_tension}. We use the tension statistics in Eq.\ (\ref{eq:T}) to quantify agreement. For the Shah et al.\ data we use the posteriors these authors provided us for comparison for the case when the cosmological background has a dark energy equation of state modeled by the CPL parametrization, and the perturbation parameters $\alpha_{M,B}$ are related by $\alpha_M=-\alpha_B$, and $\alpha_M=c_M\Omega_\Lambda/\Omega_{\Lambda 0}$.

\begin{figure}[h!]
    \centering
\includegraphics[width=\columnwidth]{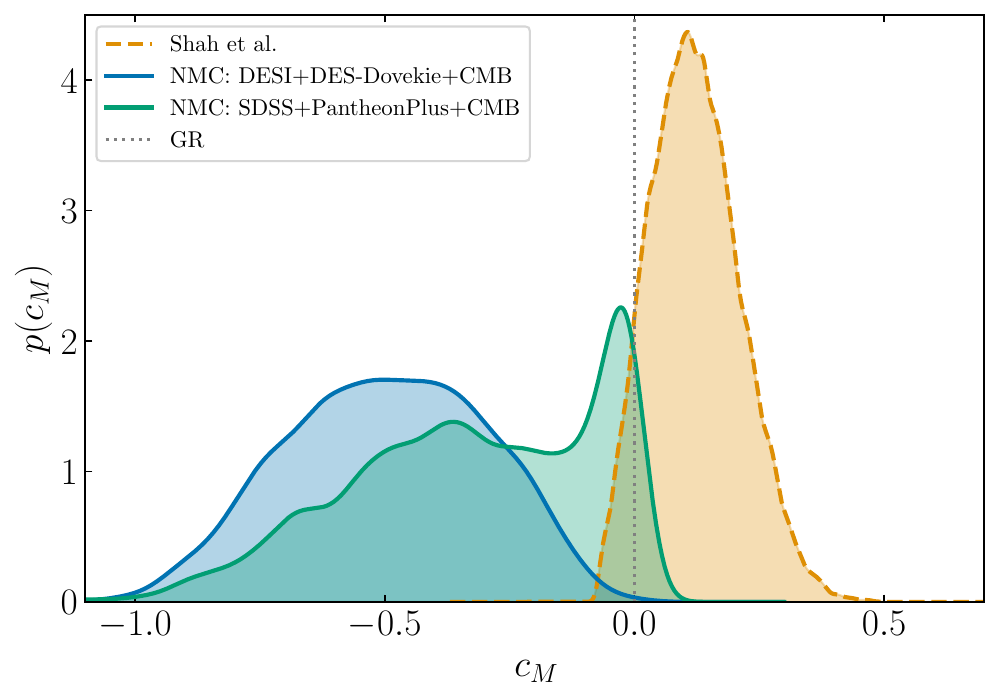}
    \caption{Comparison of $c_M$ posteriors between the NMC model and Shah et al. 
     \textit{Orange dashed}: Shah et al.\ with theoretical constraint $\alpha_M=-\alpha_B$, $c_M=0.12\pm 0.09$.
    \textit{Blue solid}:
    inferred from the NMC posterior (DESI+DES-Dovekie+CMB). \textit{Green solid}:
    inferred from the NMC posterior (SDSS+PantheonPlus+CMB).
     The vertical
    grey dot-dashed line marks GR.}
    \label{fig:GWcompare3}
\end{figure}

Fig.\ \ref{fig:GWcompare3} compares the posterior from Shah et al.\ in dashed orange, $c_M=0.12\pm 0.09$, with the one from the NMC model in solid blue (DESI+DES-Dovekie+CMB), $c_M=-0.5\pm 0.2$. Interestingly, the constraints they obtain are quite disjoint with the ones we obtain here and are suggestive of a tension between the analyses. In this case, we obtain a tension between these two posteriors of $\mathcal{T}=2.7$.

Some differences should be expected as our data combination from which the $c_M$ derived is more favorable to evolving dark energy than the data used in Shah et al.\ as both DESI BAO and DES SNe datasets display a notably stronger preference for dynamical dark energy \cite{DESI:2025zgx, DESI:2024mwx}. This would of course lead to more significant scalar field evolution in the NMC model (and thus a more sharply evolving $\alpha_M$ and a more negative $c_M$). In order to determine whether the datasets display this much tension with each other or if there are other contributing factors to the discrepancy between the inferred $c_M$ constraints, we perform a re-analysis of the NMC model deploying a similar data combination to Shah et al.\, consisting of SDSS DR16 BAO and $f\sigma_8$ measurements \cite{eBOSS:2020yzd} (where we use the BOSS, eBOSS LRG, Quasar, and Lya measurements of Table III), Planck CMB temperature and polarization power spectra \cite{Planck:2018vyg, Planck:2019nip}, Planck+ACT lensing \cite{ACT:2023kun}, and PantheonPlus SNe \cite{Scolnic:2021amr}. 

While the above is a similar analysis to that deployed in Shah et al, it is also important that we are clear about the differences. One very important factor that hinders a direct comparison between our model and the Shah et al.\ results is that their analysis of the CMB data varies the lensing parameter $A_L$, which is connected with the well-known lensing anomaly whereby the data seems to prefer an $A_L>1$ (whereas $A_L$ physically should be equal to unity). While there may be good reasons to vary this parameter in certain contexts (e.g.\ as a consistency or to conservatively marginalize over unaccounted systematics), varying this parameter is \textit{unphysical} as it reflects an arbitrary rescaling of the lensing power spectrum $C_{\ell}^{\phi \phi} \rightarrow A_L C_{\ell}^{\phi \phi}$. Consequently, when investigating specific \textit{microphysical} models of scalar field dark energy (as opposed to parametric models which have more general aims), it would not be appropriate to perform ad hoc modifications to their empirical predictions. Additionally, a smaller difference is that they include an ISW likelihood analysis, which we forgo for the present purposes as current ISW only has weak constraining power and \cite{CGG_PGF_WJW_2026} has recently shown that the cosmological evolution of the NMC model is consistent with current ISW measurements.

\begin{figure}[h!]
	\centering
	\includegraphics[width=\columnwidth]{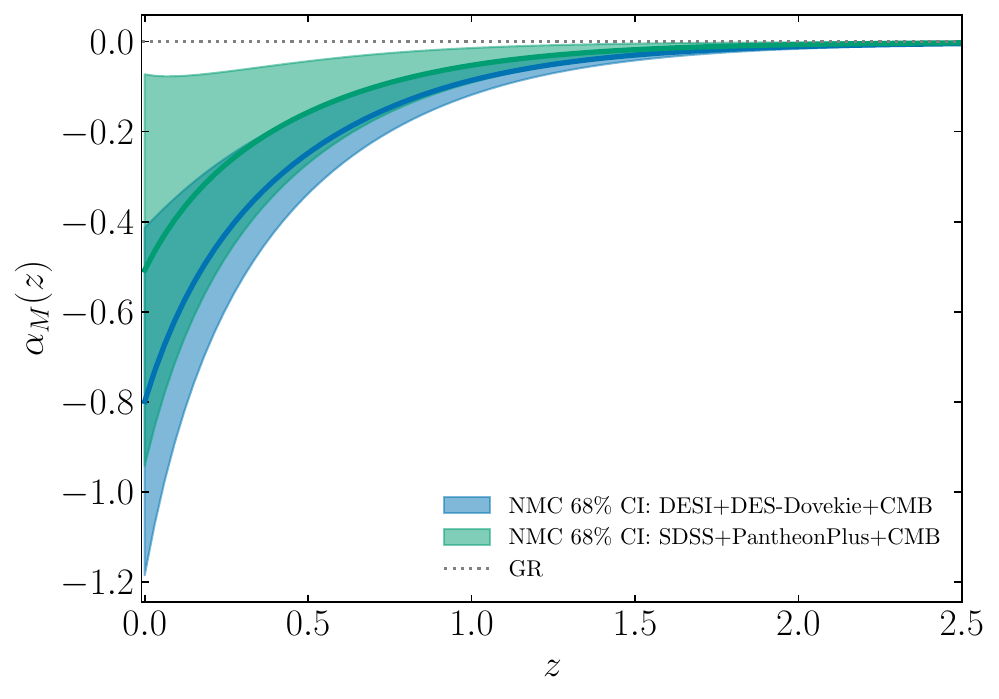}
	\caption{Posterior distribution of $\alpha_M(z)$ from the NMC quintessence model of Wolf et al.~\cite{Wolf:2025jed} determined from the combination of DESI+DES-Dovekie+CMB compared with that determined by the combination of SDSS+PantheonPlus+CMB. The bands represent the 68$\%$ CI and the solid line is the posterior median.}
	\label{fig:alphaM_overlay}
\end{figure}

Proceeding with the above analysis, we compute the theory predictions with \texttt{hi\_class} \cite{CLASS, hi_class1, hi_class2} and use \texttt{Cobaya} \cite{Cobaya, Cobaya2} to perform the MCMC sampling and obtain posterior constraints. This gives us constraints on $\alpha_M(z)$ from this alternative data combination as shown in Fig.\ \ref{fig:alphaM_overlay} (analogous to Fig.\ \ref{fig:alphaM}), and we then fit $c_M$ to these predictions as before. We notice that the old data is more consistent with the $\Lambda$CDM model, yielding less dynamical evolution in $\alpha_M$, with $\alpha_M(z=0)=-0.45^{+0.42}_{-0.50}$ and $\alpha_M(z=0)=-0.76^{+0.37}_{-0.42}$ for the older and newer data respectively. This can also been seen when extracting the dark energy effective equation of state $w_\varphi$: the newer DESI+DES-Dovekie+CMB data leads to an inference of $w_\varphi(z=0) = -0.87 \pm 0.03$, while the SDSS+PantheonPlus+CMB data leads to an inference of $w_\varphi(z=0) = -0.91 \pm 0.05$.

The result on the fitted $c_M$ is shown in solid green in Fig.\ \ref{fig:GWcompare3}, with $c_M=-0.3\pm{0.3}$, and in this case the tension with Shah et al.\ reduces to $\mathcal{T}=1.5$ and the results are more compatible as they share overlap around the GR expectation of $c_M=0$.  While there are clear differences in preferences for more dynamical evolution, the corresponding fitted $c_M$ from the NMC model for the two data sets (solid blue and green in Fig.\ \ref{fig:GWcompare3}) are consistent with each other, with $\mathcal{T}=0.6$.

In other words, the data sets are consistent, but display expected differences with respect to the inferred $c_M$ parameter when they are both consistently interpreted through the same underlying NMC model. However, there are still some striking differences with the Shah results which utilized the $\alpha_{M,B}$ model parameterization: the constraints from Shah et al.\ give a preference for the \textit{opposite} signs in $c_M$ from the ones we have derived here. Indeed, while the green posterior in Fig.\ \ref{fig:GWcompare3} shows a very slight support for positive $c_M$ values, this is simply an artifact of smoothing the MCMC samples with KDE, since no actual samples have $c_M>0$ due to a physically motivated prior that keeps $\alpha_M > 0$\footnote{This has to do with the fact that the scalar field equation of motion (EoM) is dynamically less stable for $\xi < 0$ in $F(\varphi)=1-\xi\varphi^2$. In the EoM, we have a curvature-dependent effective mass $m^2_{\rm eff} = m^2+\xi R$. Furthermore, $R\simeq 0$ in radiation domination, while $R\propto H^2$ during matter domination. If $\xi<0$, the potential will generally have a maximum during matter domination as the $R$ term will dominate the effective mass for values of $m^2$ typical for late-time scalar field dark energy (which are of order $H_0^2$), causing the field to rapidly roll and become dynamically unstable. This behavior could possibly be avoided with very large $m^2$ during matter domination, but this would lead to early rather than late-time dark energy (and indeed this has been considered as an early dark energy candidate \cite{Braglia:2020auw}). Thus, this branch does not naturally have the properties that would serve as a good model for late-time dark energy and we do not consider it, meaning that $\alpha_M < 0$ and $c_M < 0$ in all of the parameter space considered here.}.
The opposite signs on $c_M$ are physically significant: the NMC model drives $\alpha_M$ upward in the past (negative $\alpha_M$ means larger GW
amplitude, $D_L^{\rm GW}<D_L^{\rm EM}$), while the agnostic cosmological fit in Shah et al.\
prefers an $\alpha_M$ that decreases with time (positive
$\alpha_{M}$ means smaller GW
amplitude, $D_L^{\rm GW}>D_L^{\rm EM}$). Naively, we would expect the Shah et al.\ and our SDSS+PantheonPlus+CMB constraints on the NMC model's $c_M$ values to display more agreement with each other, even if they are not strictly speaking inconsistent (due to the small slice of overlapping constraints around the $\Lambda$CDM expectations). 

\begin{figure}[h!]
	\centering
	\includegraphics[width=\columnwidth]{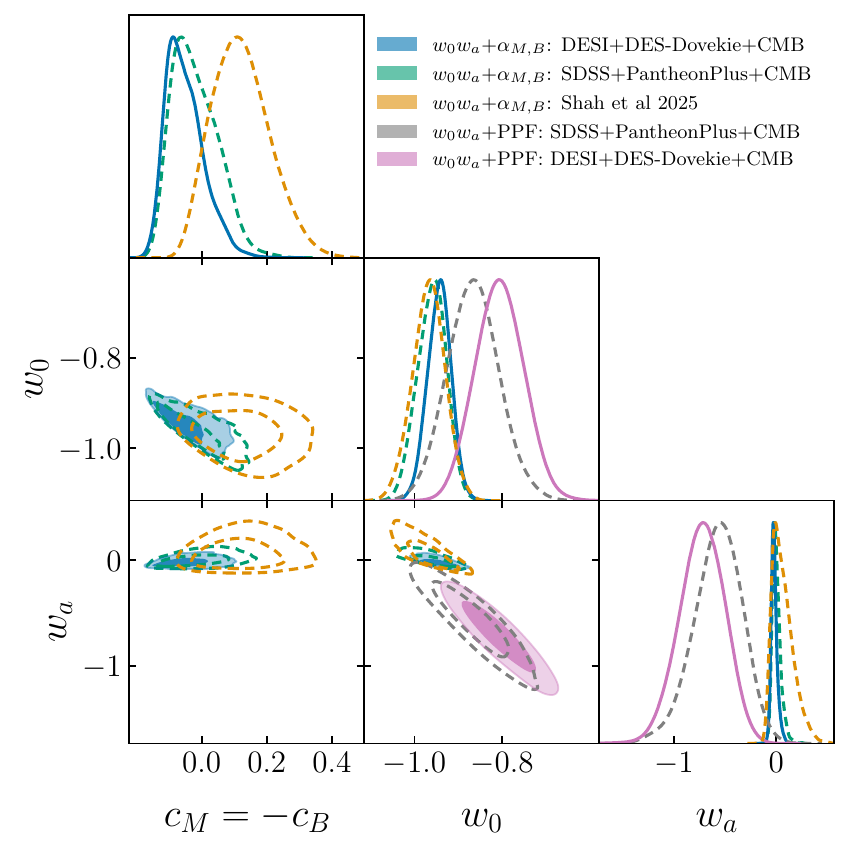}
	\caption{68\% and 95\% CI for the $w_0w_a$+$\alpha_{B,M}$ model (with the $c_M=-c_B$ theory prior) for both DESI+DES-Dovekie+CMB and SDSS+PantheonPlus+CMB datasets, in addition to the equivalent constraints for the standard $w_0w_a$+PPF cosmology for the same datasets.}
	\label{fig:triangle_c_M_constraints}
\end{figure}

To investigate further, we perform our own analysis of the $\alpha_{B,M} \propto \Omega_\Lambda$ parameterization. Using both previously mentioned data combinations (DESI+DES-Dovekie+CMB and SDSS+PantheonPlus+CMB), we constrain $c_M$ from the $\alpha_{B,M} \propto \Omega_\Lambda$ parametrization,  modeling the dark energy equation of state with the CPL parametrization $w_0w_a$ as was also done in Shah et al. The posteriors for the perturbation parameter $c_M$ and the background parameters $w_0w_a$ are depicted in Fig.\ \ref{fig:triangle_c_M_constraints}. Here, for illustrative purposes, we have also depicted the results determined from constraining the background CPL parameters in the absence of the $\alpha$'s, in which case the dark energy perturbations are evolved according to the parameterized post-Friedmann (PPF) framework \cite{Fang:2008sn}---the standard treatment of the CPL parameterization in most analyses and the default treatment in \texttt{CLASS} \cite{CLASS}.\footnote{The PPF description replaces the usual fluid density and momentum perturbation variables with a single dynamical variable that smoothly handles the phantom divide (in the fluid variables the fluid velocity can diverge as $w\rightarrow-1$), while retaining conservation of energy and the expected behavior of dark energy perturbations on both large and small scales \cite{Fang:2008sn}.} 

There are a few things to note about the results depicted in Fig.\ \ref{fig:triangle_c_M_constraints}. (1) Our results on the parametrized model $w_0w_a$+$\alpha_{B,M}$ for the SDSS+PantheonPlus+CMB analysis are visually much closer to the Shah analysis as it clusters around $c_M = 0$ and is much more tightly constrained than the NMC constraints.  We now obtain $c_M = -0.03^{+0.08}_{-0.06}$ (SDSS+PantheonPlus+CMB) yielding a tension $\mathcal{T} = 1.2$ with Shah et al. While the $c_M$ median is certainly closer to Shah, the errors are much smaller and thus the tensions between our NMC and $\alpha_{B,M}$ analyses and Shah end up being comparable. (2) Shah et al.\ results are less constraining than ours as varying the $A_L$ parameter notably widens the constraints by introducing new degeneracies. (3) There are significant differences between the constraints on the background $w_0w_a$ parameters from the usual $w_0w_a$+PPF analysis and those coming from the  $w_0w_a$+$\alpha_{B,M}$ analysis. Introducing the perturbation parameters $\alpha_{B,M}$ into the $w_0w_a$ background cosmology  impacts the inference of dark energy equation of state parameters, with the $\alpha$'s driving dark energy towards the $\Lambda$ equation of state, in contrast with the $w_0w_a$+PPF model which notably favors more significant dark energy evolution. For instance, for the DESI+DES-Dovekie+CMB dataset we obtain $(w_0,w_a)=(-0.94\pm 0.03,-0.02^{+0.02}_{-0.03})$ for the $w_0w_a$+$\alpha_{B,M}$ model and  $(w_0,w_a)=(-0.81\pm 0.06,-0.73\pm 0.22)$ for the $w_0w_a$+PPF model. This behavior was independently corroborated by the recent reanalysis of Shah et al.\ \cite{Shah:2026etz} with DESI data in the case where $\alpha_B$ and $\alpha_M$ are independent parameters.
(4) Furthermore, while we might expect that the addition of two new parameters ($c_M$ and $c_B$) to the $w_0w_a$ model would improve the fit to the cosmological data given the additional freedom, it turns out this curiously produces worse fits to the data with $\Delta \chi^2 \simeq -5$ (DESI+DES-Dovekie+CMB) and $\Delta \chi^2 \simeq -2$ (SDSS+PantheonPlus+CMB) in the favor of the $w_0w_a$+PPF model, where this $\Delta\chi^2$ is a true log-likelihood ratio from the respective datasets (not to be confused with the goodness of fit statistic defined in Eq.~(\ref{eq:chi2})).

In addition to the above, we would like to focus on a certain curiosity particular to the newer data. While all three analyses (NMC, $w_0w_a$+PPF, and $w_0w_a$+$\alpha_{B,M}$) are not technically incompatible when it comes to the older SDSS+PantheonPlus+CMB data due to this data's consistency with $\Lambda$CDM, tensions clearly increase when ones considers the DESI+DES-Dovekie+CMB data.  From the $w_0w_a$+$\alpha_{B,M}$ model we have $c_M =  -0.06^{+0.07}_{-0.05}$ whereas for the NMC model we had $c_M = -0.5\pm 0.2$. This yields a tension for $c_M$ of $\mathcal{T} = 2.2$  between the $w_0w_a$+$\alpha_{B,M}$ and NMC models. Analogously, we can quantify the tension in the $w_0w_a$ parameters by using the 2D tension metric 
\begin{equation}\label{eq:2d_tension}
\mathcal{T} = \Delta\mu^{T}
\left(C_1+C_2\right)^{-1}
\Delta\mu,
\end{equation}
where $\mu_i = (\langle w_0\rangle_i,\langle w_a\rangle_i)$ denotes the median vector in the $w_0w_a$ parameter space and $C_i$ is the corresponding covariance matrix for the $i$th posterior. And then $\mathcal{T}$ is converted to a $\sigma$ value by treating it as a $\chi^2$ statistic with 2 dof.
The tension between the $w_0w_a$+$\alpha_{B,M}$ and $w_0w_a$+PPF models yields $\mathcal{T}=9.95$, which becomes a $\simeq 2.7 \sigma$ tension. Here, we use the median to be consistent with the tension metric we defined in Eq.~(\ref{eq:T}), but the posteriors for $w_0$ and $w_a$ are very close to Gaussian so the difference between using the mean or median is negligible.

The fact that the $w_0w_a+\alpha_{B,M}$ constraints are in tension with constraints from the NMC and $w_0w_a$+PPF results is because the treatment of perturbations in the $\alpha_{B,M}\propto\Omega_{\rm \Lambda}$ case severely restricts the dark energy dynamics. At issue here is the sound speed $c_s^2$, which defines how pressure and density perturbation in the dark energy fluid respond to each other with $c_s^2 \equiv \left(\frac{\delta p}{\delta \rho}\right)$, and is required to be positive in order to avoid gradient instabilities. In the standard treatment of the $w_0w_a$+PPF model, the speed of sound of the dark energy perturbations does not factor into the analysis since the usual fluid variables are replaced in favor of alternative variables that smoothly handle the phantom divide crossing.
On the other hand, within microphysical models like quintessence or the NMC, there is a series of remarkable cancellations in the various quantities that determine the $c_s^2$ expression, which simply results in $c_s^2=1$ and ensures that such microphysical models are free of gradient instabilities. However, in parameterizing the perturbations using $\alpha_{M,B}$ while parameterizing the background using $w_0w_a$, one effectively decouples the scalar field and cosmological background terms that together produce these delicate cancellations. Thus, one can very easily end up with a set of parameter choices \{$c_M, c_B, w_0, w_a$\} that generates a gradient instability through a negative $c_s^2$ (as can be seen analytically in Appendix \ref{app:cs2}). In other words, both the NMC and standard CPL models are naturally protected from gradient instabilities over a much wider range of dynamical dark energy behaviors, whereas the presence of these instabilities in $\alpha_{M,B}$ model automatically excludes much of the equation of state parameter space that deviates from $\Lambda$.

\begin{figure}[h!]
	\centering
	\includegraphics[width=\columnwidth]{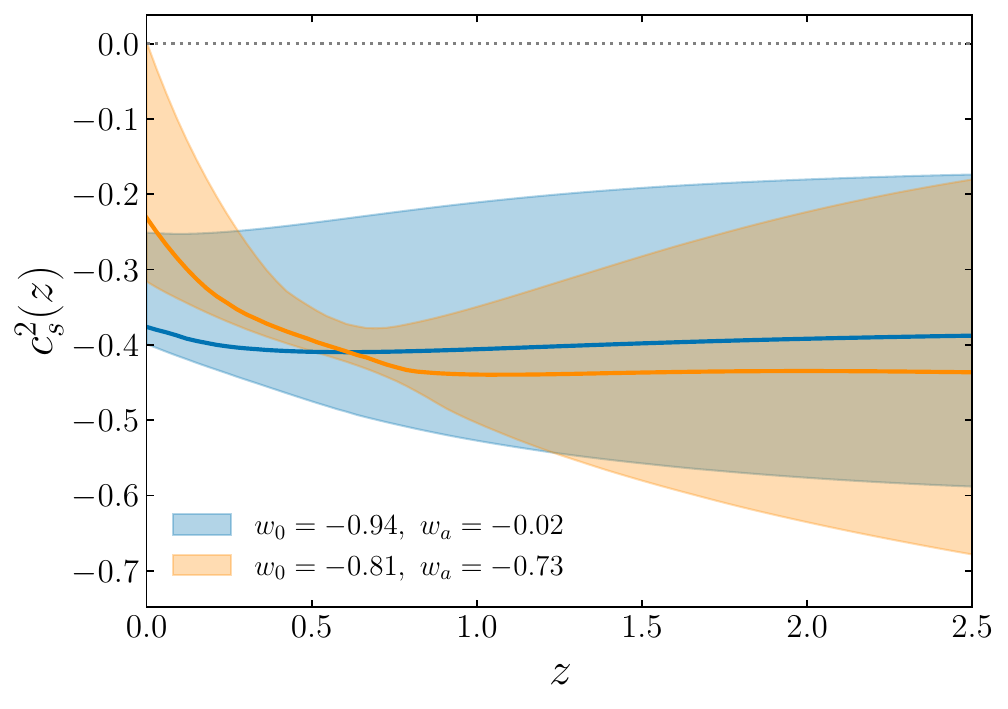}
	\caption{Median (solid line) and 68\% CI (shaded region) for the evolution of the sound speed $c_s^2(z)$ according to the $w_0w_a$+$\alpha_{M,B}$ model, for the $c_M = -c_B$ values derived from the $\alpha_M$ of the NMC model using the DESI+DES-Dovekie+CMB data (c.f.\ Fig.\ \ref{fig:alphaM_overlay}) for two representative choices of $w_0$ and $w_a$. Perturbative stability requires $c_s^2>0$, yet here virtually all of the parameter space is deeply unstable across the redshift ranges where dark energy becomes a significant fraction of the energy density. }
	\label{fig:cs2}
\end{figure}

To illustrate the extent to which the parameter space is constrained by gradient instabilities, we sample the $\Omega_m$ and $c_M$ values derived from the $\alpha_M$ constraints of the microphysical NMC model in Fig.\ \ref{fig:alphaM_overlay} (for the DESI+DES-Dovekie+CMB data combination) and calculate the $c_s^2(z)$ evolution within the CPL+$\alpha_{M,B}$ model (according to Eq.\ (\ref{eq:cs2})) over the redshift range $z\in [0.0,2.5]$ for two sets of representative $(w_0,w_a)$ values.\footnote{For these purposes, we also specify the value of the kineticity $\alpha_K = c_K \Omega_\Lambda/\Omega_{\Lambda0}$, arbitrarily setting $c_K\simeq 0.7$. While this leads to generally larger values than what the true value of $\alpha_K$ would be in the actual NMC model, smaller values only make the unstable behavior worse (i.e.~$c_s^2$ becomes even more negative than it already is). Indeed, when keeping everything else constant ($w_0$, $w_a$, $c_M$, $c_B$), one finds that the form of the evolution of $c_s^2(z)$ can change substantially depending on the choice of $c_K$. However, even raising $c_K$ by multiple orders of magnitude does not change the result that this parameter space is plagued with gradient instabilities for the combination of $c_M=-c_B$ and $w_0w_a$ values considered here.} For these, we choose the means resulting from the posteriors of the $w_0w_a$+PPF model $(w_0,w_a)=(-0.81,-0.73)$ as well as the $w_0w_a+\alpha_{B,M}$ model $(w_0,w_a)=(-0.94,-0.02)$, both from the DESI+DES-Dovekie+CMB data constraints. The evolution of $c_s^2(z)$ is depicted in Fig.\ \ref{fig:cs2}, where it is evident that the entire parameter space defined by the combination of $c_M$ values derived from the NMC model and the chosen $w_0w_a$ values is dominated by gradient instabilities with $c_s^2 < 0$ over nearly the entire redshift range. These parameter combinations will thus all be automatically rejected when performing cosmological inference. 

Furthermore, these gradient instabilities are not just numerical artifacts as sometimes arise due to machine precision; rather, these values are large in magnitude and negative over recent redshifts where the dark energy density has become a significant fraction of the total energy density. In other words, the values for $c_M$ that we infer from analyzing NMC dynamics are simply not accessible within the $\alpha_{B,M}\propto\Omega_{\rm \Lambda}$ model for a massive portion of the $w_0w_a$ parameter space that best describes expansion history data.

In summary, the parameter space of  $(w_0,w_a,c_{M})$ within the $w_0w_a$+$\alpha_{M,B}$ model is significantly constrained due to gradient instabilities that are not present in either the $w_0w_a$+PPF model nor quintessence-like scalar field models like NMC. The simple parametrization $\alpha_{M,B}\propto \Omega_\Lambda$ restricts the dark energy dynamics in a way that is incompatible with the rich $\alpha_M(z)$ (and fully stable) evolution of the NMC model, leading to conflicting constraints on $c_M$. 

\begin{figure}[h!]
	\centering
	\includegraphics[width=\columnwidth]{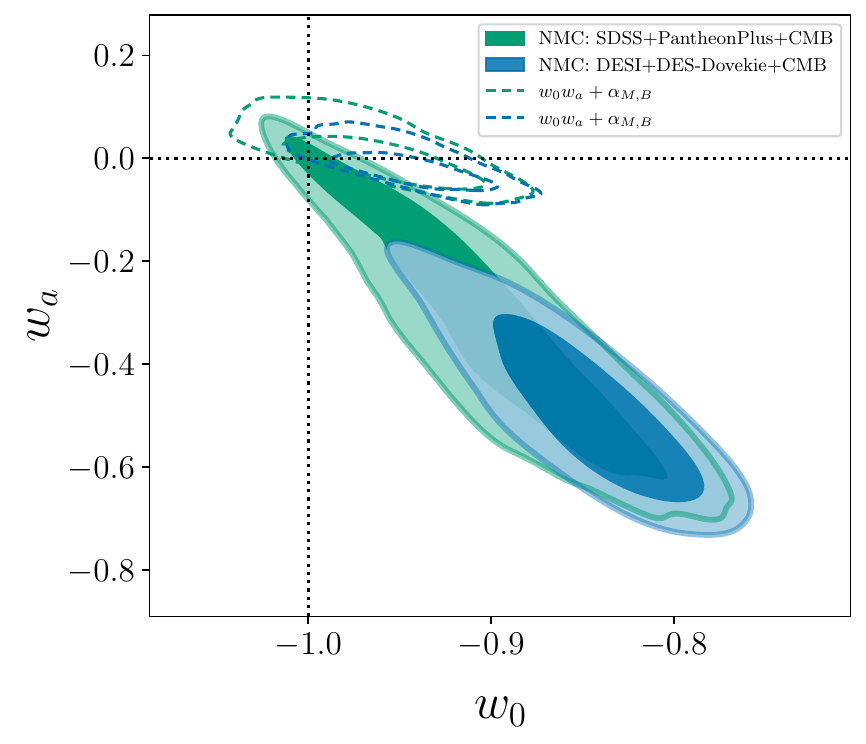}
	\caption{Posterior distributions of $w_0$ and $w_a$ (68\% and 95\% CI) from the $w_0w_a$+$\alpha_{M,B}$ (dashed) compared with the inferred posterior distributions on the parameters determined from fitting the CPL model to the observable predictions of the NMC model (solid) for both the SDSS+PantheonPlus+CMB (green) and DESI+DES-Dovekie+CMB (blue) data. As before, the results using SDSS+PantheonPlus dataset are not inconsistent due to the shared overlap around the $\Lambda$ prediction, but when utilizing the newer DESI+DES-Dovekie dataset we begin to see a tension between the parametric and microphysical models. We trace this tension back to the presence of instabilities in the $w_0w_a$+$\alpha_{M,B}$ model when $c_M < 0$ and $w_0w_a$ displays sufficiently dynamical thawing behavior (c.f.~Fig.~\ref{fig:cs2} for an illustration of the gradient instabilities in the $w_0w_a$+$\alpha_{M,B}$ model for representative $w_0w_a$ values).  }
	\label{fig:NMC_w0wa_fit}
\end{figure}

As alluded to above, these gradient instabilities also bias the inference for the background evolution of the equation of state, as is evident in the tension between the $w_0w_a$ constraints favored by $w_0w_a$+PPF and $w_0w_a$+$\alpha_{M,B}$ analyses. 
This has further implications for the NMC model as well; when one fits $w_0w_a$ to the background evolution of the NMC model as has been done in \cite{Wolf:2025jed, Wolf:2024stt, CGG_PGF_WJW_2026}\footnote{The fitting technique here involves fitting the $w_0w_a$ parameters to both the observable predictions of the NMC model along with the exact redshift bins and errors of the cosmological surveys relevant to these observables (e.g.\ $D_V(z)$ in the case of the BAO measurements). See \cite{Wolf:2025jed, Wolf:2024stt, CGG_PGF_WJW_2026, Wolf:2025jlc} for more details.}, one finds that the fitted $w_0w_a$ share substantial overlap with those determined from the $w_0w_a$+PPF model while being disjoint from the $w_0w_a$ determined by the $w_0w_a$+$\alpha_{M,B}$ model. This is depicted in Fig.~\ref{fig:NMC_w0wa_fit}, where we see that the inferred $w_0w_a$ share no overlap with the $w_0w_a$+$\alpha_{M,B}$ constraints when using the DESI+DES-Dovekie+CMB data (while we again see a small overlap with the SDSS+PantheonPlus+CMB specifically in the $\Lambda$ compatible region of parameter space). In addition to the $\mathcal{T} = 2.2$ tension in $c_M$, we have from Eq.~(\ref{eq:2d_tension}) that the NMC model and the $w_0w_a$+$\alpha_{M,B}$ model constraints on $w_0w_a$ have a $\simeq 3.6 \sigma$ tension with each other. 

In other words, while we would naively expect both of the $c_M$ and $w_0w_a$ inferred from the $w_0w_a$+$\alpha_{M,B}$ model to align with those inferred from the NMC model, the parametrized model turns out to be qualitatively and quantitatively misleading as the condition of gradient instabilities in the $w_0w_a$+$\alpha_{M,B}$ model automatically excludes the phenomenology characterized by the combination of both the running Planck mass (captured in $c_M$) and the background evolution of the dark energy equation of state thawing (captured in $w_0w_a$) that we infer from the fully stable dynamical evolution of the NMC model. Furthermore, this pathological behavior clearly becomes more pronounced as one moves away from the $\Lambda$ limits of the parameter space.

While parameterizations can be useful in certain contexts, these results do call into question the utility of using parameterizations of $\alpha_{M,B}$ in cosmological inference, as they fail to comprehensively capture the perturbative and background phenomenology in even very simple microphysical models that we would naively expect them to cover like the NMC model. 
We stress that this limitation is specific to parametrizations of the perturbation sector subject to a stability prior, and does not undermine the background-level GW-propagation parametrizations of Sec.\ \ref{sec:GW_agreement}. The latter are direct fits to the distance ratio $D_L^{\rm GW}/D_L^{\rm EM}$, which carry no perturbative stability requirement, and therefore track the NMC prediction to percent-level precision (as seen in Fig.\ \ref{fig:ratio}). By contrast, the $\alpha_{B,M}\propto \Omega_\Lambda$ parametrization couples the background and perturbation dynamics and is subject to the gradient-stability condition $c_s^2>0$, which excludes precisely the dynamical dark-energy region occupied by the NMC model. This conclusion agrees with the recent reanalysis of Shah et al.\ with DESI data \cite{Shah:2026etz}. It is this stability-driven implicit prior---not the use of a parametrization per se---that makes the $w_0w_a+\alpha_{B,M}\propto \Omega_\Lambda$ model a poor proxy for microphysical models.

A possible way around this problem is to perform model-independent reconstructions of dark energy parameters, which would still decouple background and perturbation dynamics but may offer more freedom than $\alpha_{M,B}\propto \Omega_\Lambda$ models, which would in turn potentially expand the viable parameter space for $c_M$. 
Data-driven reconstructions have been developed at the
background level, e.g.\ \cite{Holsclaw_2010, Seikel_2012, Nesseris_2012, Dialektopoulos_2022, Mitra_2024, DESI:2025fii}, where it is possible to reconstruct the expansion history and the
dark energy equation of state $w(z)$ directly from the data without assuming a
functional form, and closely related agnostic methods have been applied
specifically to the GW luminosity distance and its associated propagation
(friction) history \cite{Belgacem:2019zzu, Yang:2021qge}. The same techniques can be
extended to the perturbation sector \cite{Espejo:2018hxa,Pogosian:2021mcs,Raveri_2023}.
In this sense, agnostic reconstructions offer a natural alternative to the $\alpha$
parametrizations.

We caution, however, that this flexibility comes at the cost of a direct
microphysical interpretation: a purely data-driven reconstruction---particularly
one based on a flexible neural network or Gaussian process---does not map onto a
specific Lagrangian, and therefore cannot by itself confirm or exclude a concrete
scenario such as the NMC model studied here. Rather, the two approaches are
complementary. A stable microphysical model like the NMC theory provides a
concrete, falsifiable prediction for the joint background-and-perturbation
phenomenology---including a definite, self-consistent $c_s^2$ and $c_M(z)$---whereas
agnostic reconstructions provide a way to test whether the data favor any departure
from $\Lambda$CDM without imposing the restrictive priors implicit in the
$\alpha_{M,B}\propto\Omega_\Lambda$ family. 
Hybrid approaches that reconstruct functions within certain classes of theoretical models can be also applied \cite{Dialektopoulos:2023jam, marcos2026reconstructionshorndeskisubclassesgaussian}. 
Comparing an agnostic reconstruction directly against the NMC prediction, rather than against a parametrized $\alpha$
template, is thus arguably the cleaner consistency test, and we leave such a
dedicated implementation to future work.

\section{Discussion and Outlook}
\label{sec:discussion}

In this work we have asked whether the non-minimally coupled (NMC) dark energy
model of Wolf et al.~\cite{Wolf:2025jed}, which is preferred over
$\Lambda$CDM by current background cosmological data, makes gravitational-wave
(GW) propagation predictions that are consistent with existing standard-siren
constraints, and how these predictions compare with the agnostic
parametrizations commonly used in both GW and large-scale-structure analyses. To
do so, we projected the $\alpha_M(z)$ posterior of the NMC model---derived from
DESI-DR2 BAO, Planck and ACT CMB, and DES-Dovekie SNe data---onto the GW
observable $D_L^{\rm GW}/D_L^{\rm EM}$, fitted the standard $c_M$ and
$(\Xi_0,n)$ parametrizations to it, and compared the resulting values against the
GWTC-5 standard-siren constraints and the cosmological constraints
of Shah et al.~\cite{Shah:2025vnt}. We also quantified the bias that a nonzero NMC
$\alpha_M$ would induce on the value of $H_0$ inferred from standard sirens.

The NMC model predicts a GW friction $\alpha_M(z)<0$ at low redshift, with
$\alpha_M(z{=}0)=-0.76^{+0.37}_{-0.42}$, decaying to zero by $z\sim2$. This
corresponds to a GW/EM luminosity-distance ratio below unity at all redshifts,
dropping from one at low $z$ to $\simeq0.88$ by $z\sim1.5$, so that GW sources
appear systematically closer than their EM counterparts would predict. Projected
onto the standard parametrizations this yields $c_M=-0.5\pm 0.2$ and
$(\Xi_0,n)=(0.88\pm0.05,\,3.2\pm0.3)$. We find the two-parameter $(\Xi_0,n)$ template
to reproduce the NMC distance ratio better than the one-parameter $c_M$ model, because the fixed
$\Omega_\Lambda(z)$ redshift dependence built into $c_M$ cannot capture the richer background evolution of the NMC model.

These NMC predictions for $c_M$ and
$(\Xi_0,n)$ are fully consistent with the current GWTC-5 constraints.
Using our tension metric we find $\mathcal{T}=0.5$ ($0.08$) for $c_M$ and $\mathcal{T}=0.7$
($0.7$) for $\Xi_0$ under the narrow (wide) $H_0$ prior. This consistency should
not be over-interpreted, however: the current GW constraints have $1\sigma$
uncertainties of order unity, comfortably encompassing both the NMC prediction ($c_M=-0.5$, $\Xi_0=0.88$)
and GR ($c_M=0$, $\Xi_0=1$). The GW data are therefore not yet able to
distinguish the NMC model from GR, and the agreement represents an encouraging
but not yet stringent cross-probe validation.

If the NMC signal is present but a standard-siren
analysis using the $c_M$ parametrization assumes GR ($c_M=0$), the inferred Hubble constant is biased high.
Reading the best-fit $H_0$ along the GWTC-5 $H_0$--$c_M$ degeneracy at
$c_M=-0.5$, we find a bias of $\Delta H_0=2.8$~km/s/Mpc under the
narrow prior and $\Delta H_0=5.9$~km/s/Mpc under the wide prior would be expected. This bias is comparable in magnitude to the Hubble tension itself,
and has the right sign to alleviate it: correctly accounting for the NMC friction
moves the GW-inferred $H_0$ downward, toward the Planck value. For the $(\Xi_0,n)$ parametrization, the bias essentially negligible under the narrow prior ($\Delta H_0=0.6$~km/s/Mpc) but leads to a shift of $\Delta H_0=5.8$ km/s/Mpc for the wide prior. 

The decisive test of the NMC GW signature will come from improved standard-siren
statistics. Forecast for current LVK detectors show that the NMC model and GR will not be distinguished with statistical significance in the coming years. Instead, next-decade detectors in the ground and space, such as Cosmic Explorer, Einstein Telescope and LISA, will change this, with forecasts predicting percent-level constraints on modified gravity parameters, sufficient to distinguish the NMC model and GR at high significance. 

In contrast, the agnostic $c_M$ constraint obtained by Shah et al.\ from
CMB+LSS+SNe data using the $\alpha_M\propto\Omega_\Lambda$ parametrization is in
 tension with the NMC prediction, with $\mathcal{T}=2.7$, and---crucially---prefers the opposite sign, $c_M>0$. We traced this discrepancy to a structural limitation of
the parametrized approach: for a $w_0w_a$ background with
$\alpha_M=-\alpha_B=c_M\Omega_\Lambda/\Omega_{\Lambda 0}$, a negative $c_M$
drives the scalar-field sound speed negative ($c_s^2<0$), triggering a gradient
instability across the redshifts where dark energy is dynamically relevant. The
parametrized model therefore cannot access the region of parameter space occupied
by the NMC model and returns support for $c_M$ close to 0 or positive. This is a concrete manifestation of
the implicit-prior problem: agnostic $\alpha$ parametrizations do not faithfully
represent even relatively simple, fully stable microphysical models such as the
NMC one, and the apparent tension is an artifact of the parametrization rather
than a genuine physical disagreement.

Taken together, these results establish that
current GW data are consistent with the NMC model, that its GW signature is a
genuine and falsifiable prediction distinct from GR, and that the apparent
tension with agnostic cosmological constraints reflects the shortcomings of the
$\alpha\propto\Omega_\Lambda$ parametrization rather than a failure of the model.
Using such parametrizations as a proxy for
microphysical dark energy when interpreting either GW or LSS data needs to be analyzed carefully.

Several directions would sharpen the analysis presented here. Because the GWTC-5
modified-propagation constraints assume a flat-$\Lambda$CDM expansion history
with fixed $\Omega_m=0.3065$, whereas the NMC model predicts an evolving
dark-energy fraction and $\Omega_m=0.311$, a fully consistent comparison
ultimately calls for a joint parameter estimation in which the expansion history
and the GW friction are fitted simultaneously within the NMC model. It would also
be valuable to compare the NMC prediction directly against non-parametric
reconstructions of $\alpha_M(z)$ or
$D_L^{\rm GW}/D_L^{\rm EM}(z)$, in order to avoid unphysical parameter restrictions as the one found for the $\alpha_M\propto \Omega_\Lambda(z)$ parametrization. We
leave these dedicated analyses for future work.


\begin{acknowledgments}
We thank N.\ Shah and  J.\ Noller for providing their posteriors for comparison in this paper. M.L.\ was supported by Fondecyt Iniciación grant No.\ 11250105.
\end{acknowledgments}

\appendix

\section{Fitting to the NMC model}

\label{app:fitting}
We fit both the $c_M$ and $(\Xi_0, n)$ parametrizations to the NMC posterior 
by performing a per-sample non-linear least-squares minimization. 
For each of the $N=2500$ posterior samples at a given redshift, we minimize

\begin{equation}\label{eq:chi2_app}
    \chi^2_i(\vec{p}) = \sum_j \frac{\left(R_i(z_j) - R_{\rm model}(z_j;\vec{p})\right)^2}{\sigma_r^2(z_j)},
\end{equation}
where $R_i(z_j) = D_L^{\rm GW}/D_L^{\rm EM}$ is the distance ratio computed from 
the $i$-th NMC sample via Eq.~(\ref{eq:ratio}), $R_{\rm model}$ is the 
parametrized model either using Eq.\ (\ref{eq:cM}) or (\ref{eq:Xi0n}). In the case of the $c_M$ parametrization, the distance ratio depends on $\Omega_m$, which is obtained from the NMC samples corresponding to the a given sampled line in Fig.\ \ref{fig:alphaM} with $\Omega_{m,i}$, consistently with the sample's own distance-ratio curve $R_i(z_j)$. Then, the distance ratio residual is normalized by the fractional $68\%$ uncertainty $\sigma_r(z_j) = \sigma_R(z_j)/\bar{R}(z_j)$ from the NMC $R_i$ distribution at each redshift $z_j$. Here,  $\bar{R}$ is the median from the NMC samples at fixed $z_j$, and  $\sigma_R$ is the $68\%$CI across the $N$ samples (see footnote \ref{foot:sigma} for how to handle non-gaussian distributions).  
We use the fractional uncertainty as the weighting since the NMC posterior  exhibits a roughly constant relative error of $\sim 5\%$ across all redshifts, whereas the absolute uncertainty varies significantly and downweighs high redshift meaningful information. We restrict the fit to $z \leq 2.5$, as the NMC posterior converges to $R\to 1$ at high redshift where the signal carries little information about the model parameters. 

The minimization of $\chi^2_i$ yields one best-fit $\hat{\vec{p}}_{i}$ per sample line, and the posterior distribution  $p(\vec{p})$ is constructed from the collection $\{\hat{\vec{p}}_{i}\}$. This minimization procedure is performed using the python function \texttt{scipy.optimize.curve\_fit}. For the $c_M$ parametrization we obtain a median and $68\%$CI of $c_M=-0.52^{+0.22}_{-0.24}$, whereas for the $(\Xi_0,n)$ parametrization we obtain $\Xi_0=0.878^{+0.049}_{-0.048}$ and $n=3.21^{+0.34}_{-0.33}$. In Eqs.\ (\ref{cM_value}) and (\ref{xi0n_values}) we quote these results with one less significant digit.

We emphasize that the fitting procedure could have been chosen differently and similar results would have been obtained. For instance, if we had fitted $c_M$ directly to the $\alpha_M$ samples in Fig.\ \ref{fig:alphaM} (using an analogous $\chi^2$ formula as in Eq.\ (\ref{eq:chi2_app})) we would have obtained a median and $68\%$CI of $c_M=-0.49^{+0.21}_{-0.21}$. Similarly, since the fractional error $\sigma_r$ is quite constant for the redshift range considered, we could have also fitted the parametrized models by simply minimizing residuals (i.e.\ using the $\chi^2$ expression in Eq.\ (\ref{eq:chi2_app}) without the denominator that weights by fractional error) and we would have obtained $c_M=-0.50^{+0.21}_{-0.23}$. Notice that these two results are the same as the one quoted in Eq.\ (\ref{cM_value}) from fitting the distance ratio with Eq.\ (\ref{eq:chi2_app}), to one significant digit.

\section{Sound speed constraints}\label{app:cs2}
In order to ensure the stability of cosmological perturbations, the squared sound speed of the scalar field (dark energy) perturbations must be positive. For models where dark energy perturbations are parametrized by $\alpha_{K,M,B,T}$ \cite{Bellini:2014fua} the expression for the squared sound speed is given by 
\begin{align}\label{eq:cs2}
    c_s^2=&\frac{1}{D}\left\{  (2-\alpha_B)\left( \alpha_M+\frac{\alpha_B}{2}-\frac{\dot{H}}{H^2}\right) \right.\nonumber\\
    & \left. -\frac{(\rho_m+p_m)}{H^2M^2} +\frac{\dot{\alpha}_B}{H}\right\}
\end{align}
when GWs propagate at the speed of light (i.e.\ have a parameter $\alpha_T=0$). Here, $D=\alpha_K+3\alpha_B^2/2$, and $M$ is the running Planck mass. This expression depends on background quantities such as $H$ and $(\rho_m,p_m)$, as well as the perturbation parameters $(\alpha_M,\alpha_B,\alpha_K)$. Stability conditions require $c_s^2>0$, which will translate into constraints in the parameter space for the background and perturbations. 

On the contrary, for a given Horndeski model, the scalar field dynamics determines both the background evolution and the perturbation parameters, and stability can be translated directly onto constraints on the fundamental coupling constants in the action. Explicit expressions for $c_s^2$ in Horndeski can be found in \cite{Bellini:2014fua}. In the NMC model, one can check that the condition $\alpha_B=-\alpha_M$ is satisfied and the sound speed takes a constant value 
$c_s^2=1$, hence always satisfying the stability constraint. 

Nonetheless, in a general parametrized model with $w_{\rm DE}=w_0+(1-a)w_a$ and $\alpha_M=-\alpha_B=c_M\Omega_{\Lambda}(a)/\Omega_{\Lambda 0}$ one finds that:
\begin{align}\label{eq:cs2_z0}
    c_s^2(z=0)=& \frac{1}{D}\left\{ \frac{c_M}{2}[5+3w_0(1+\Omega_{m})]+\frac{c_M^2}{2}\left(\frac{1+\Omega_{m}}{1-\Omega_{m}}\right)\nonumber \right.\\
    &\left. +3[1+w_0(1-\Omega_{m})-\Omega_{m}]\right\}
\end{align}
for the value of the sound speed today. In the NMC model we have that $\alpha_K=\dot{\varphi}^2>0$ whereas in the EFTDE $\alpha_K=c_K\Omega_{\Lambda}(z)/\Omega_{\Lambda 0}$ with $c_K$ typically chosen as a small positive value. Therefore, we always consider $D>0$ and the sign of $c_s^2$ is determined by the numerator in Eq.\ (\ref{eq:cs2_z0}). 
If $c_M<0$ then the numerator can become negative and break the stability condition $c_s^2>0$. Indeed, in the parameter region where $-1.5<w_0<-0.5$ and $0<\Omega_{m}<1$ then a $c_M$ value of -0.5 (as preferred by the NMC model) would indeed yield a $c_s^2(z=0)<0$ in most of the region. This is why the parametrized approach cannot reproduce correctly the NMC model and instead provides posteriors with support for $c_M\approx 0$ or $c_M> 0$.


\bibliography{references}

\end{document}